\documentclass[a4paper,11pt]{article}
\pdfoutput=1 

\usepackage{jheppub} 

\usepackage[utf8]{inputenc} 

\usepackage[T1]{fontenc} 

\usepackage{amsmath, amssymb, amsthm, graphicx, epsfig, fancyhdr,epsfig, slashed}

\usepackage{tikzsymbols}
\usepackage{natbib}
\usepackage{float}
\usepackage{tikz,xcolor,hyperref}
\usepackage{bm}
\usepackage{url}
\usepackage{braket}
\usepackage{mathtools}
\usepackage{placeins}
\usepackage{booktabs}
\usepackage{graphicx}
\usepackage{bbold}
\usepackage[normalem]{ulem}

\makeatletter \gdef\@fpheader{}\makeatother

\newcommand{\beq}{\begin{equation}}
\newcommand{\eeq}{\end{equation}}

\newcommand{\be}{\begin{equation}}
\newcommand{\ee}{\end{equation}}
\newcommand{\bea}{\begin{eqnarray}}
\newcommand{\eea}{\end{eqnarray}}
\newcommand{\bal}{\begin{aligned}}
\newcommand{\eal}{\end{aligned}}

\newcommand{\nn}{\nonumber}

\title{ \boldmath Constraining dark matter from strong phase transitions in a $U(1)_{L_{\mu}-L_{\tau}}$ model: Implications for neutrino masses and muon $g-2$}

\author[a,b]{Sandhya Choubey,}
\author[c]{Sarif Khan,}
\author[a,b]{Marco Merchand}
\author[a,b]{and Sampsa Vihonen}

\affiliation[a]{KTH Royal Institute of Technology, 
Department of Physics, SE-10691 Stockholm, Sweden}
\affiliation[b]{The Oskar Klein Centre for Cosmoparticle Physics, AlbaNova University Centre, SE-10691 Stockholm, Sweden}
\affiliation[c]{
    Institut f\"{u}r Theoretische Physik, 
    Georg-August-Universit\"{a}t G\"{o}ttingen,
    Friedrich-Hund-Platz 1,
    37077 G\"{o}ttingen, Germany
}

\emailAdd{sandhya.choubey@gmail.com}
\emailAdd{sarif.khan@uni-goettingen.de}
\emailAdd{marcomm@kth.se}
\emailAdd{vihonen@kth.se}

\abstract{ In this paper, we study a non-minimal gauged $U(1)_{L_{\mu}-L_{\tau}}$ model, where we add two complex singlet scalars, three right-handed Majorana neutrinos (RHN), and a vector-like dark fermion to the Standard Model (SM), all non-trivially charged under the extra gauge symmetry. The model offers an easy resolution to the muon $(g-2)$ anomaly, which fixes the scale of spontaneous symmetry breaking. Furthermore, the two-zero minor structure in the RHN mass matrix provides successful predictions for neutrino oscillation parameters, including the Dirac phase.
The extended scalar sector can easily induce first-order phase transitions. We identify all possible phase transition patterns in the three-dimensional field space. 
We quantify the associated gravitational waves from the sound wave source and demonstrate that the signatures can be observed in future space-based experiments. We find that strong first-order phase transitions require large values of scalar quartic couplings  which constrain the scalar dark matter (DM) relic density to a maximum of $10^{-2}$ and $10^{-5}$ 
when we consider the DM direct detection bound. 
Nonetheless, the model successfully explains the DM relic density via contribution from the vector-like dark fermion. We show the allowed range of the model parameters that can address all the beyond SM issues targeted in this study.

}

\begin{document}
\maketitle
\flushbottom

\section{Introduction}

The standard model of particle physics (SM) is arguably the most successful theory to describe the physics of elementary particles. Despite its agreement with experimental observations \cite{Workman:2022ynf}, the SM alone is unable to account for several well-established properties of the Universe. The deficiencies of the SM include (i) the observed neutrino masses and their oscillation parameters, (ii) the particle identity of the dark matter (DM) and (iii) a consistent explanation of the present day baryon asymmetry. In order to accommodate the aforementioned issues simultaneously, one must consider beyond the SM (BSM) theories that may include new fundamental interactions, particles and symmetries. 

 One of the simplest extensions to the SM can be obtained by introducing an Abelian gauge symmetry, under which the second and third lepton generations are non-trivially charged. This extension is commonly referred to as the $U(1)_{L_{\mu} -L_{\tau}}$ extension \cite{Foot:1990mn,He:1990pn,He:1991qd}. Under the new gauge symmetry, leptonic particles of flavor $\mu$ have charge $+1$ while those of flavor $\tau$ have charge $-1$, hence the name $L_{\mu} -L_{\tau}$. For each lepton generation, we include a corresponding Majorana right-handed neutrino (RHN), which is a singlet under the SM. In this way the smallness of the neutrino masses is explained via the type-I seesaw mechanism. The $U(1)_{L_{\mu} -L_{\tau}}$ gauge symmetry is free from anomalies and is assumed to be spontaneously broken by the vacuum expectation value (vev) of a complex singlet scalar, providing a mass to the associated $Z'$ gauge boson.  These assumptions give rise to a very particular RHN neutrino mass structure, which is commonly known as the \textit{two-zero minor} structure \cite{Araki:2012ip,Crivellin:2015lwa,Asai:2017ryy, Asai:2018ocx, Asai:2019ciz} since the diagonal $22$ and $33$ entries vanish due to the extra symmetry. The model with this particle content has been dubbed the \textit{minimal} gauged $U(1)_{L_{\mu} -L_{\tau}}$ in Refs.~\cite{Asai:2017ryy,Asai:2018ocx,Asai:2019ciz} and has been studied in the context of the baryon asymmetry via leptogenesis \cite{Asai:2020qax,Granelli:2023egb}, particle DM phenomenology \cite{Baek:2008nz,Biswas:2016yan,Biswas:2016yjr} and $(g-2)_{\mu}$ anomaly \cite{Heeck:2011wj,Baek:2001kca,Ma:2001md,Borah:2021jzu}, to name the most common.

Particularly interesting is the case in which more than one deficiency of the SM can be simultaneously accounted for in the same scenario. This is the case for the model introduced in Refs.~\cite{Biswas:2016yan, Biswas:2016yjr}, where the \textit{minimal} gauged  $U(1)_{L_{\mu} -L_{\tau}}$ was responsible for providing a DM particle candidate consistent with experimental constraints, neutrino masses and neutrino oscillation parameters. The model was also able to solve the $(g-2)_{\mu}$ anomaly~\cite{Muong-2:2021ojo, Aoyama:2020ynm}. 
On the other hand, incorporating viable leptogenesis while maintaining the correct $(g-2)_{\mu}$ value is unfeasible, as the latter requires spontaneous symmetry breaking at the scale $\mathcal{O}(1-100)$ GeV while the former needs very large RHN masses\footnote{The compatibility of leptogenesis  with the $(g-2)_{\mu}$ anomaly can be  attained by invoking a non-minimal version of the $U(1)_{L_{\mu} -L_{\tau}}$ model with two complex scalars with non-zero vevs, see Refs.~\cite{Eijima:2023yiw,Borah:2021mri}.}, {\it i.e.}, $M_i \gtrsim \mathcal{O}(10^9)$ GeV \cite{Davidson:2002qv}.

The discovery of gravitational waves (GWs) by the LIGO/Virgo/KAGRA Collaboration~\cite{LIGOScientific:2016aoc,LIGOScientific:2016sjg,LIGOScientific:2017bnn,LIGOScientific:2017vox,LIGOScientific:2017ycc,LIGOScientific:2018mvr,LIGOScientific:2020ibl} from inspiral binary systems and the potential detection by other experiments such as the Pulsar Timing Array (PTA) collaborations \cite{NANOGrav:2023gor,EPTA:2023sfo} and the future Laser Interferometer Space Antenna (LISA)~\cite{LISA:2017pwj} have sparked a new gold rush in the characterization of all possible signals. Of notable relevance for particle physics are those coming from early universe first-order phase transitions (PTs) which are tightly connected to the structure of the scalar potential and highly influenced by the size of any Higgs-portal interactions. Furthermore, first-order PTs can provide the required out-of-equilibrium condition necessary for electroweak (EW) baryogenesis~\cite{Morrissey:2012db}. It is well known that strong PTs prefer rather large Higgs-portal interactions as they strengthen the potential barrier between two potential minima~\cite{Lewicki:2024xan,Curtin:2014jma,Beniwal:2017eik,Beniwal:2018hyi}. In contrast, a successful explanation of the DM in the \textit{minimal} gauged $U(1)_{L_{\mu} -L_{\tau}}$ model \cite{Biswas:2016yan, Biswas:2016yjr} requires a sufficiently small Higgs-portal interaction as otherwise direct detection experiments \cite{LZ:2022lsv} rule out the model. The incompatibility between particle DM and the appearance of strong PTs in the \textit{minimal} scenario has prompted us to consider a slight modification of the original theory.  

In the present paper we study a non-minimal gauged $U(1)_{L_{\mu} -L_{\tau}}$ by adding to the \textit{minimal} scenario an extra complex singlet scalar with vanishing vev and a vector-like fermion. These extra fields are singlets under the SM gauge group and their non-zero charges under $U(1)_{L_{\mu} -L_{\tau}}$ are parameterized by arbitrary constants $q_x$ and $q_{\psi}$, respectively. This choice is not only generic but it also makes it possible to regulate the direct couplings between the extra fields and the $Z'$ gauge boson, which control the production of DM via the freeze-out mechanism.  

As one would expect, the model achieves the solution to the $(g-2)_{\mu}$ anomaly and successfully predicts the neutrino oscillation parameters. In this context, we undertake a major revision of the neutrino sector by using the most recent \texttt{NuFit} data~\cite{Esteban:2020cvm} and provide the scale of the fundamental parameters for which the predicted neutrino oscillation parameters fall within their experimental limits. 

The extended scalar sector of our model warrants a careful investigation of its finite temperature potential. We undertake such endeavour keeping track of all the dynamical fields which renders the effective potential a 3-dimensional function in field-space. We then perform a systematic scan of the free parameters and find the regions for which strong PTs can be produced. Along the way we examine the multiple transition patterns that can arise and investigate useful correlations between the fundamental scalar couplings and the strength of the PT. After having uncovered the relevant parameter space of PTs we assess the predicted GW signals coming from the sound wave source. This allows to further restrict the parameter space by considering only benchmark points which predict detectable signals at LISA.

As the last step of our work, we study the two-component DM phenomenology in which the dark fermion mass and $U(1)_{L_{\mu} -L_{\tau}}$ charge are the only remaining free parameters. This allows us to easily find solutions which abide by DM relic abundance and direct detection bounds.

In summary, our extended gauged $U(1)_{L_{\mu} -L_{\tau}}$ scenario can account for: (i) the $(g-2)_{\mu}$ anomaly, (ii) neutrino oscillation observables, (iii) strong PTs and associated GWs and provides (iv) a two-component particle DM consistent with all current limits.

The paper is structured as follows. In section~\ref{sec:model}, we introduce the model and set the notation writing down the general Lagrangian and the free parameters in the model. In section~\ref{sec:g-2}, we review the solution to the $(g-2)_{\mu}$ anomaly. We then proceed to review the status of neutrino oscillation parameters in section~\ref{sec:neutrinos}, discussing the numerical fitting procedure in detail. In section~\ref{sec:finiteT}, we present the one-loop effective potential along with the finite temperature contributions, showing the parameter space which is consistent with first-order PTs and classify all the possible patterns of field-space dynamics. Section~\ref{sec:GWs} is dedicated to examine the associated GW signals. It is shown that that the predicted GW signals can be probed in the future GW experiments. In section~\ref{section_DM} we investigate DM and review the most stringent constraints and provide the viable parameter space. Following that, in section~\ref{sec:GW_DM}, we provide the viable parameter space that is consistent with all observations and display the most interesting correlations in Lagrangian parameters. We finally provide our conclusions and outlook in section~\ref{sec:conclusions}.

\section{The model}
\label{sec:model}

In this section we present the details of our non-minimal gauged $U(1)_{L_{\mu} -L_{\tau}}$ model. For completeness, the particle content and charges are presented in Table~\eqref{tab2}. Notice that the only SM particles with non-vanishing charges are the $\mu$ and $\tau$ charged leptons. The model includes three RHN Majorana neutrinos $N_R^i$ ($i=e, \mu, \tau $) and a vector-like fermion $\psi$ which is one of the DM components. The scalar sector contains the SM Higgs doublet $\Phi$ and two extra complex scalars which we denote by $\Phi'$ and $\Phi_{DM}$. The former is responsible for spontaneous symmetry breaking of the gauged $U(1)_{L_{\mu} -L_{\tau}}$ and provides the mass to the associated $Z'$ boson, while the latter does not acquire a vev and comprises  the second DM component. It is worthwhile to stress that all new degrees of freedom are neutral under the SM gauge group.
\begin{table}[!t]
\centering
\resizebox{\textwidth}{!}{%
\begin{tabular}{||c|c|c|c|c||}
\hline
\hline
\begin{tabular}{c}
    Gauge\\
    Group\\ 
    \hline
    $U(1)_{L_{\mu} - L_{\tau}}$\\ 
    
\end{tabular}
&
\begin{tabular}{c}
    \multicolumn{1}{c}{Baryonic Fields}\\ 
    \hline
    $(Q^{i}_{L}, u^{i}_{R}, d^{i}_{R})$\\ 
    \hline
    $0$ \\ 
\end{tabular}
&
\begin{tabular}{c}
    \multicolumn{1}{c}{Vector-like}\\ 
    \hline
    $\psi_L$, $\psi_R$\\ 
    \hline
    $q_{\psi}$ \\ 
\end{tabular}
&
\begin{tabular}{c|c|c}
    \multicolumn{3}{c}{Lepton Fields}\\ 
    \hline
    $(L_{L}^{e}, e_{R}, N_{R}^{e})$ & $(L_{L}^{\mu}, \mu_{R},
    N_{R}^{\mu})$ & $(L_{L}^{\tau}, \tau_{R}, N_{R}^{\tau})$\\ 
    \hline
    $0$ & $1$ & $-1$\\ 
\end{tabular}
&
\begin{tabular}{c|c|c}
    \multicolumn{3}{c}{Scalar Fields}\\
    \hline
    $\Phi$ & $\Phi'$ & $\Phi_{DM}$ \\
    \hline
    $0$ & $1$ & $q_{DM}$\\
\end{tabular}\\
\hline
\hline
\end{tabular}}
\caption{Particle content and their corresponding
charges under the gauge group $U(1)_{L_{\mu} - L_{\tau}}$.}
\label{tab2}
\end{table}

The complete Lagrangian for the model 
can be expressed as
\begin{eqnarray}
\mathcal{L}&=&\mathcal{L}_{\text{SM}} + \mathcal{L}_{\text{scalar}} + \mathcal{L}_{\psi}  +\mathcal{L}_{Z'} +\mathcal{L}_{N}  \,,
\label{lag}
\end{eqnarray}
where $\mathcal{L}_{SM}$ denotes the full SM Lagrangian but does not include the scalar sector. The scalar sector is given by the second term, $\mathcal{L}_{\text{scalar}}$, and will be introduced in the next subsection. The third term contains the vector-like dark fermion, namely $\mathcal{L}_{\psi} = \bar{\psi} \left( i \gamma^{\mu} D_{\mu} - m_{\psi}\right) \psi$,
where the covariant derivative contains the non-trivial charges:
\begin{equation}
    D_{\mu} X \equiv \left( \partial_{\mu}  - i g_{Z'}  Q(X)Z'_{\mu}  \right) X, 
\end{equation}
where $X=\psi, N_R^e,N_R^{\mu},N_R^{\tau}, \Phi'$, $\Phi_{DM}$ and $Q(X)= ( q_{\psi}, 0, 1, -1, 1, q_{DM})$ are the respective charges. Here $g_Z'$ is the new gauge coupling constant. We keep the charge $Q(X)$ as a free parameter for the 
fermion and scalar DM defined as $q_{\psi}$ and $q_{DM}$, respectively. The physical mass of the $U(1)_{L_{\mu} - L_{\tau}}$ gauge boson after the 
spontaneous symmetry breaking takes the form,
\begin{equation}
    m_{Z'} = g_{Z'}  v',
\end{equation}
where $v'$ is the vev of $\Phi'$. The dynamics of the extra gauge boson is encoded in its kinetic term via $\mathcal{L}_{Z'} = \frac{1}{4} F_{\mu \tau}^{\alpha \beta} {F_{\mu \tau}}_{\alpha \beta}$. In this work we ignore any possible kinetic mixing with the photon. The effects of the kinetic mixing have been studied elsewhere~\cite{Foldenauer:2024cdp, Escudero:2019gzq,Croon:2020lrf}.

Following the particle charge assignments from table~\eqref{tab2}, the neutrino sector can be written as 
\begin{eqnarray}
\mathcal{L}_{N}&=&
\sum_{i=e,\,\mu,\,\tau}\frac{i}{2}\bar{N_i}\gamma^{\mu}D_{\mu} N_{i} 
-\dfrac{1}{2}\,M_{ee}\,\bar{N_e^{c}}N_{e}
-\dfrac{1}{2}\,M_{\mu \tau}\,(\bar{N_{\mu}^{c}}N_{\tau}
+\bar{N_{\tau}^{c}}N_{\mu})  \nn \\ &&
-\dfrac{1}{2}\,h_{e \mu}(\bar{N_{e}^{c}}N_{\mu} 
+\bar{N_{\mu}^{c}}N_{e})\Phi'^\dagger
- \dfrac{1}{2}\,h_{e \tau}(\bar{N_{e}^{c}}N_{\tau} 
+ \bar{N_{\tau}^{c}}N_{e})\Phi'
\nn \\ &&
-\sum_{i=e,\,\mu,\,\tau} y_{i} \bar{L_{i}}
\tilde {\Phi} N_{i} +h.c.
\label{lagN}
\end{eqnarray}
This piece of the Lagrangian gives rise to the celebrated \textit{two-zero minor} structure \cite{Araki:2012ip,Crivellin:2015lwa,Asai:2017ryy, Asai:2018ocx, Asai:2019ciz}. We will review the neutrino sector, in full detail, in section~\ref{sec:neutrinos}.

\subsection{Tree level scalar potential}

The scalar sector of our model is characterized by its tree-level potential, which we write in the most generic form:
\begin{align}
  V_0(\Phi, \Phi',\Phi_{\text{DM}}) =& V_{\text{SM}}(\Phi) + \mu_{DM}^{2} \Phi_{DM}^{\dagger} \Phi_{DM} 
+\lambda_{DM} (\Phi_{DM}^{\dagger} \Phi_{DM})^{2}
\nonumber \\
& + \mu_{\phi'}^{2} \Phi'^{\dagger} \Phi' 
+ \lambda_{\phi'} (\Phi'^{\dagger} \Phi')^{2} \nonumber \\
&+ \lambda_{D\phi}(\Phi_{DM}^{\dagger} \Phi_{DM})
(\Phi^{\dagger} \Phi) 
+\lambda_{D\phi'}(\Phi_{DM}^{\dagger} \Phi_{DM})(\Phi'^{\dagger} \Phi') \nonumber \\
&+ \lambda_{\phi \phi'}(\Phi'^{\dagger} \Phi') (\Phi^{\dagger} \Phi),
\label{V0}
\end{align}
where $ V_{\text{SM}}(\Phi) = -\mu_{\phi}^2 \Phi^{\dagger} \Phi + \lambda_{\phi} (\Phi^{\dagger} \Phi)^2$ is the SM Higgs potential.

Since the Higgs doublet $\Phi$ and the $U(1)_{L_{\mu} - L_{\tau}}$ singlet $\Phi'$ get vev at zero temperature, we write them as
\begin{eqnarray}
\Phi=
\begin{pmatrix}
G^+ \\
\dfrac{v + \phi + i \ G_{\phi}}{\sqrt{2}}
\end{pmatrix}\,,
\,\,\,\,\,\,\,\,\,
\Phi^{\prime}=
\begin{pmatrix}
\dfrac{v^{\prime} + \phi^{\prime} + i \ G_{\phi^{\prime}}}{\sqrt{2}}
\end{pmatrix}\,,
\label{phih}
\end{eqnarray}
where $G^{\pm}$, $G_{\phi}$ and $G_{\phi'}$ are the Goldstone bosons which are absorbed into the longitudinal components of the  $W^{\pm}$, $Z$ and $Z'$ gauge bosons, respectively. The DM candidate is contained in the extra $U(1)_{L_{\mu} - L_{\tau}}$ complex singlet which we parameterize as 
\begin{equation}
    \Phi_{DM} =\dfrac{ \chi e^{i \alpha}  }{\sqrt{2}},
\end{equation}
where $\alpha$ is an arbitrary phase. This choice is to emphasize that after spontaneous symmetry breaking there are two scalar degrees of freedom with degenerate mass, $m_{DM}$.

The initial total of $9$ free parameters in \eqref{V0} are reduced to only $7$ by means of the phenomenological constraints $v=246$ GeV and $m_{\phi}=125$ GeV.  We choose as free parameters the set 
\begin{equation}
\text{Scalar sector parameters} = \{ \theta, \ v',  \ m_{\phi'}, \ m_{DM}, \ \lambda_{D\phi}, \ \lambda_{D\phi'}, \ \lambda_{DM} \}. \label{eq:scalar_params}
\end{equation}
Making use of the minimization conditions and mass matrix diagonalization, the dependent parameters are the mass squared parameters,
\begin{equation}
\begin{split}
     \mu^2_{\phi} &=  \frac{1}{2}( 2 v^2 \lambda_{\phi}   + v'^2 \lambda_{\phi \phi'} ), \\
     \mu^2_{\phi'} &=  \frac{1}{2}( - 2 v'^2 \lambda_{\phi'}   - v^2 \lambda_{\phi \phi'} ) , \\
     \mu^2_{DM} &=  m_{DM}^2-\frac{1}{2}(  v^2 \lambda_{D \phi} + v'^2 \lambda_{D\phi'}),
\label{mphidm}
\end{split}
\end{equation}
and also the quartic couplings
\begin{equation}
\begin{split}
\lambda_{\phi} &= \frac{1}{2 v^2}(m_{\phi}^2\cos^2\theta+m_{\phi'}^2\sin^2\theta) \, , \\
\lambda_{\phi'} &= \frac{1}{2v'^2}(m_{\phi}^2\sin^2\theta + m_{\phi'}^2\cos^2\theta) \, , \\
\lambda_{\phi \phi'} &= \frac{1}{v v'}((m_{\phi'}^2-m_{\phi}^2)\sin\theta\cos\theta ) \, .
\label{quartic-couplings}
\end{split}
\end{equation}

Before continuing to the next section, we provide a full counting of the free model parameters. As we have shown, the scalar sector provides us with $7$ parameters, which are listed in Eqn.~\eqref{eq:scalar_params}. The neutrino sector is described by the Yukawa couplings in Eqn.~\eqref{lagN}, which gives us 
\begin{equation}
\text{neutrino sector parameters} = \{ M_{ee}, M_{\mu \tau}, h_{e\mu}, h_{e\tau}, y_e, y_{\mu}, y_{\tau} \}.
\end{equation}
Additionally, one needs to set the vector-like fermion charge $q_{\psi}$ and mass $m_{\psi}$, the new gauge coupling constant $g_{Z'}$ and the DM scalar charge $q_x$. Therefore, the model is fully specified with a total of $18$ free parameters. As we proceed to the following sections, each phenomenological observable will fix the range of a subset of the free model parameters. We will demonstrate that the full parameter space can by narrowed down by the aforementioned considerations.


\section{The muon $(g-2)$ anomaly}
\label{sec:g-2}

The $(g - 2)_{\mu}$ value can be calculated very accurately within the SM. The SM prediction for $(g - 2)_{\mu}$ is $a_{\mu} (\text{SM}) = 116591810 (43) \times 10^{-11} (0.37 \ \text{ppm})$~\cite{Aoyama:2020ynm} and it includes corrections from electrodynamics, hadronic vacuum polarization, hadronic light by light as well as EW processes. At the same time, advancements in the experimental methodologies have led to a very precise measurement of $(g - 2)_\mu$. The most notable contributions have been carried out at Brookhaven (BNL~E821)~\cite{Muong-2:2006rrc}, CERN~\cite{Charpak:1962zz, Charpak:1965zz, Combley:1974tw, CERN-Mainz-Daresbury:1978ccd} and Fermilab~\cite{Muong-2:2021ojo}. The measured value differs from the SM prediction by~\cite{Muong-2:2021ojo, Aoyama:2020ynm}\begin{equation}
    \Delta a_{\mu}  =  a_{\mu} (\text{exp}) - a_{\mu} (\text{SM})= (251 \pm 59) \times 10^{-11},
\end{equation}
which corresponds to $4.2\,\sigma$ confidence level (CL) discrepancy.

It is noteworthy to highlight that recent progress in the lattice computation deviates from the experimental measurements of the hadronic vacuum polarization by $2.1\sigma$, which diminishes the significance of the $\Delta a_{\mu}$ discrepancy to $1.5\sigma$~\cite{Borsanyi:2020mff}. Additionally, the new measurements from the CMD-3 experiments suggest a less-significant disagreement between the experimental observations and the theoretical predictions, albeit with a remaining $2.4\sigma$ discrepancy~\cite{CMD-3:2023alj}. The new measurements from CMD-3 introduce a conflicting perspective with prior measurements of $e^{+}e^{-} \rightarrow \pi^{+}\pi^{-}$ conducted by both the same and different experiments~\cite{Davier:2017zfy, Keshavarzi:2018mgv, Colangelo:2018mtw, Davier:2019can}.

In the present work, we calculate the contribution to $(g-2)_{\mu}$ in the model. The contribution arises from the one-loop diagram mediated by the extra gauge boson, giving 
\begin{eqnarray}
    \label{delta_a_mu}
    \Delta a_{\mu} = \frac{g^2_{Z^{\prime}}}{8 \pi^{2}}
    \int_{0}^{1} dx \frac{2 x (1-x)^{2}}{(1-x)^{2} + r x}\,,
\end{eqnarray}
where $r = \frac{m^2_{\mu}}{m^2_{Z^{\prime}}}$ and $m_{\mu}$ is the muon
mass. Taking $m_{Z^{\prime}} = 100$~MeV and $g_{Z^{\prime}} = 9 \times 10^{-4}$, for example, leads to the deviation $\Delta a_{\mu} = 226 \times 10^{-11}$, which is within the experimental range specified in Eqn.~(\ref{delta_a_mu}). The allowed region in the $g_{Z^{\prime}} - m_{Z^{\prime}}$ is discussed in detail, in the context of DM, in section~\ref{section_DM}. It is shown that the model can explain the muon $(g-2)_{\mu}$ anomaly in the case the anomaly survives the future measurements on $(g-2)_{\mu}$. In contrast, the exclusion of $(g-2)_{\mu}$ can also be accommodated in the present work as a large region of the parameter space predicts a negligible contribution to $(g-2)_{\mu}$. We further note that the resolution to the $(g-2)_{\mu}$ can be potentially ruled out by a high precision observation of the luminosity from white dwarfs cooling proceeding from $Z'$ decays into neutrinos~\cite{Foldenauer:2024cdp}. However, this bound can be evaded by introducing tree-level kinetic mixing which cancels the kinetic mixing generated by the loop diagrams.

\section{Neutrino mass}
\label{sec:neutrinos}

The neutrino masses in this model arise from the Lagrangian presented in Eqn.~\eqref{lagN}. The model has two kinds of mass matrices, namely the Majorana mass matrix consisting of right-handed neutrinos and the Dirac mass matrix consisting of both left- and right-handed neutrinos. The Majorana mass matrix is given by
\begin{eqnarray}
\mathcal{M}_{R} = \left(\begin{array}{ccc}
M_{ee} ~~&~~ \dfrac{ v^{\prime}}{\sqrt{2}} h_{e \mu}
~~&~~\dfrac{v^{\prime}}{\sqrt{2}} h_{e \tau} \\
~~&~~\\
\dfrac{v^{\prime}}{\sqrt{2}} h_{e \mu} ~~&~~ 0
~~&~~ M_{\mu \tau} \,e^{i\xi}\\
~~&~~\\
\dfrac{v^{\prime}}{\sqrt{2}} h_{e \tau} ~~&
~~ M_{\mu \tau}\,e^{i\xi} ~~&~~ 0 \\
\end{array}\right),
\end{eqnarray}
whereas we write the Dirac mass matrix, in diagonal form without loss of generality
\begin{eqnarray}
M_{D} = \left(\begin{array}{ccc}
y_e ~~&~~ 0 ~~&~~ 0 \\
~~&~~\\
0 ~~&~~ y_{\mu} ~~&~~ 0 \\
~~&~~\\
0 ~~&~~ 0 ~~&~~ y_{\tau} \\
\end{array}\right) \,.
\label{md}
\end{eqnarray}
On the other hand, the full neutrino mass matrix reads
\begin{eqnarray}
M = \left(\begin{array}{cc}
0 & M_D \\
M^T_D & M_R
\end{array}\right) \,\,.
\label{mtot}
\end{eqnarray}

The neutrino mass matrix in Eqn.~\eqref{mtot} can be diagonalised in different limits of $M_{D}$ and $M_{R}$, for example, the pseudo-Dirac limit $M_{D} \gg M_{R}$ and the seesaw limit $M_{D} \ll M_{R}$. In the present work, we focus on the seesaw limit $M_{D} \ll M_{R}$, in which case the eigenvalues take the form
\begin{eqnarray}
m_{\nu}&\simeq&-M_D\,M^{-1}_R M^T_D\,, 
\label{activemass}\\
m_N &\simeq& M_R\,,
\label{sterilemass}
\end{eqnarray} 
where $m_{\nu}$ represents the light neutrino mass matrix and $m_{N}$ represents the heavy neutrino mass matrix, respectively. 

The light neutrino mass matrix $m_\nu$ in Eqn.~\eqref{activemass} can be diagonalised to obtain the three neutrino mass eigenstates $m_1$, $m_2$ and $m_3$,
\begin{eqnarray}
    m^{\nu}_{d} = 
    diag(m_1,m_2,m_3)
    = U_{PMNS}\, m_{\nu}\, U^{T}_{PMNS}\,,
\end{eqnarray}
where $U_{PMNS}$ is the well-known Pontecorvo-Maki-Nakagawa-Sakata matrix,
\begin{eqnarray}
\label{UPMNS}
U_{PMNS} = 
    \begin{pmatrix}
       c_{12} c_{13} & s_{12} c_{13}  & s_{13} e^{- i \delta_{CP}} \\
       -s_{12} c_{23} 
       - c_{12} s_{23} s_{13} e^{i \delta_{CP}}
       & c_{12} c_{23} - s_{12} s_{23} s_{13} e^{i \delta_{CP}}
       & s_{23} c_{13} \\
       s_{12} s_{23} - c_{12} c_{23} s_{13} e^{i \delta_{CP}}
       & - c_{12} s_{23} - s_{12} c_{23} s_{13} e^{i \delta_{CP}}
       & c_{23} c_{13}
    \end{pmatrix},
\end{eqnarray}
and $c_{ij} = \cos\theta_{ij}$, $s_{ij} = \sin\theta_{ij}$ for $i, j = 1, 2, 3$. The neutrino mixing angles $\theta_{12}$, $\theta_{13}$ and $\theta_{23}$ can be derived directly from $U_{PMNS}$, as shown in appendix~\ref{app:neutrino_parameters}, whereas the $CP$ phase $\delta_{CP}$ is obtained from the Jarlskog invariant. The remaining neutrino mixing parameters $\Delta m_{21}^2$ and $\Delta m_{31}^2$ are determined from $m_1$, $m_2$ and $m_3$ as $\Delta m_{ij}^2 \equiv m_i^2 - m_j^2$. Without loss of generality, we limit our study to the normal ordering of the neutrino masses, where $m_1 < m_2 <m_3$.

We adopt the following numerical approach for the neutrino sector in this model. We generate random values for the model parameters $M_{ee}$, $M_{\mu\tau}$, $h_{e\mu}$, $h_{e\tau}$, $\xi$, $f_e$, $f_\mu$ and $f_\tau$ and use $\chi^2$ minimization to locate model parameter values that satisfy the 3$\sigma$ CL constraints for the neutrino oscillation parameters $\theta_{12}$, $\theta_{13}$, $\Delta m_{21}^2$ and $\Delta m_{31}^2$. In order to show the full predictivity of the model, $\theta_{23}$ and $\delta_{CP}$ are allowed to have values beyond their 3$\sigma$ CL ranges. The 3$\sigma$ CL ranges are adopted from \texttt{NuFit 5.2}~\cite{Esteban:2020cvm}, focusing on the values that do not account for the atmospheric neutrino data from SuperKamiokande. The $\chi^2$ function that is to be minimized is given as 
\begin{equation}
\begin{split}
    \chi^2 &= \frac{(\theta_{12}^{\rm test} - \theta_{12}^{\rm true})^2}{\sigma_{\theta_{12}}^2} + \frac{(\theta_{13}^{\rm test} - \theta_{13}^{\rm true})^2}{\sigma_{\theta_{13}}^2} + \frac{(\theta_{23}^{\rm test} - \theta_{23}^{\rm true})^2}{\sigma_{\theta_{23}}^2} + \frac{(\delta_{CP}^{\rm test} - \delta_{CP}^{\rm true})^2}{\sigma_{\delta_{CP}}^2}\\ &+ \frac{((\Delta m_{21}^2)^{\rm test} - (\Delta m_{21}^2)^{\rm true})^2}{\sigma_{\Delta m_{21}^2}^2} + \frac{((\Delta m_{31}^2)^{\rm test} - (\Delta m_{31}^2)^{\rm true})^2}{\sigma_{\Delta m_{31}^2}^2},
\end{split}
\end{equation}
where $\theta_{12}^{\rm true}$, $\theta_{13}^{\rm true}$, $\theta_{23}^{\rm true}$, $\delta_{CP}^{\rm true}$, $(\Delta m_{21}^2)^{\rm true}$ and $(\Delta m_{31}^2)^{\rm true}$ correspond to the central values of the neutrino oscillation parameters and $\sigma_{\theta_{12}}$, $\sigma_{\theta_{13}}$, $\sigma_{\theta_{23}}$, $\sigma_{\delta_{CP}}$, $\sigma_{\Delta m_{21}^2}$ and $\sigma_{\Delta m_{31}^2}$ are their associated standard deviations. To ensure that the available parameter space for $\theta_{23}$ and $\delta_{CP}$ is covered comprehensively, we accept $\chi^2$ values as large as 500. At the same time, we ensure that the experimental constraints are satisfied for all neutrino oscillation parameters except $\theta_{23}$ and $\delta_{CP}$. With this approach, we scan model parameters for $v' = 111$~GeV to obtain 10,000 points that satisfy the neutrino constraints by 3$\sigma$ CL. We then exclude the data points that are not in agreement with radiative muon decay constraints from MEG~\cite{MEGII:2023ltw,Morisi:2024yxi}, leaving 9551 valid data points. 
\begin{figure}[t]
    \centering
        \includegraphics[width=0.70\textwidth]{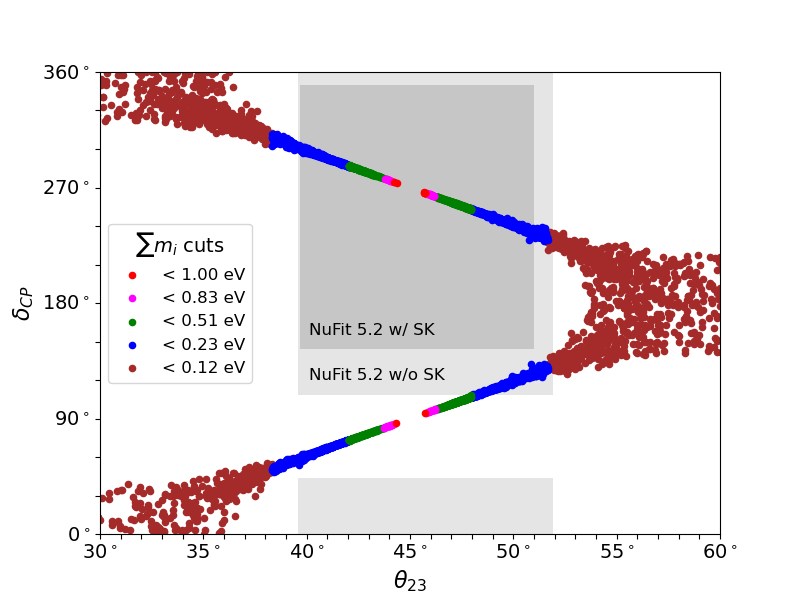}
        \caption{Predicted values for neutrino mixing parameters $\theta_{23}$ and $\delta_{CP}$. Each point satisfies the NuFit 5.2 constraints at 3$\sigma$ CL for the neutrino mixing parameters that are not shown in the figure.}
    \label{Neutrino_mixing_1}
\end{figure}

Fig.~\ref{Neutrino_mixing_1} displays the scatter plot for the neutrino oscillation parameters $\theta_{23}$ and $\delta_{CP}$ with five different cuts for the sum of light neutrino masses $\sum_i m_i$. The experimental constraints are shown for $\theta_{23}$ and $\delta_{CP}$ at 3$\sigma$ CL with the grey rectangles, which include both the constraints that take into account the SuperKamiokande (SK) atmospheric data and those that do not. The data points are marked by the different colors to show whether they satisfy the cuts $\sum_i m_i < $ 0.12~eV (brown markers), 0.23~eV (blue markers), 0.51~eV (green markers), 0.83~eV (magenta markers) and 1.00~eV (red markers), respectively. The displayed $\sum_i m_i$ cuts are chosen according to the different cosmological data sets\footnote{The cosmological bound on the sum of neutrino masses can go as low as $\sum_i m_i <$ 0.072~eV, as was found in the recent analysis of CMB and DESI~\cite{DESI:2024mwx}. The most stringent constraints apply to the $\Lambda$CDM scenario where no new physics beyond SM is included. With the introduction of new physics, the upper bound on $\sum_i m_i$ can be relaxed.} discussed in Ref.~\cite{RoyChoudhury:2018gay}. Fig.~\ref{Neutrino_mixing_1} indicates that a small number of points satisfy both the most stringent bound on neutrino masses, $\sum_i m_i < 0.12$~eV, and the constraints on neutrino oscillation parameters when the SK data is not taken into account. When we consider a higher limit on the sum, $\sum_i m_i < 0.23$~eV, a greater number of points that satisfy the constraints becomes available. We see that in this case the model predicts $\theta_{23}$ and $\delta_{CP}$ values that satisfy the constraints without SK data as well.

\begin{figure}
    \centering        \includegraphics[width=1.00\textwidth]{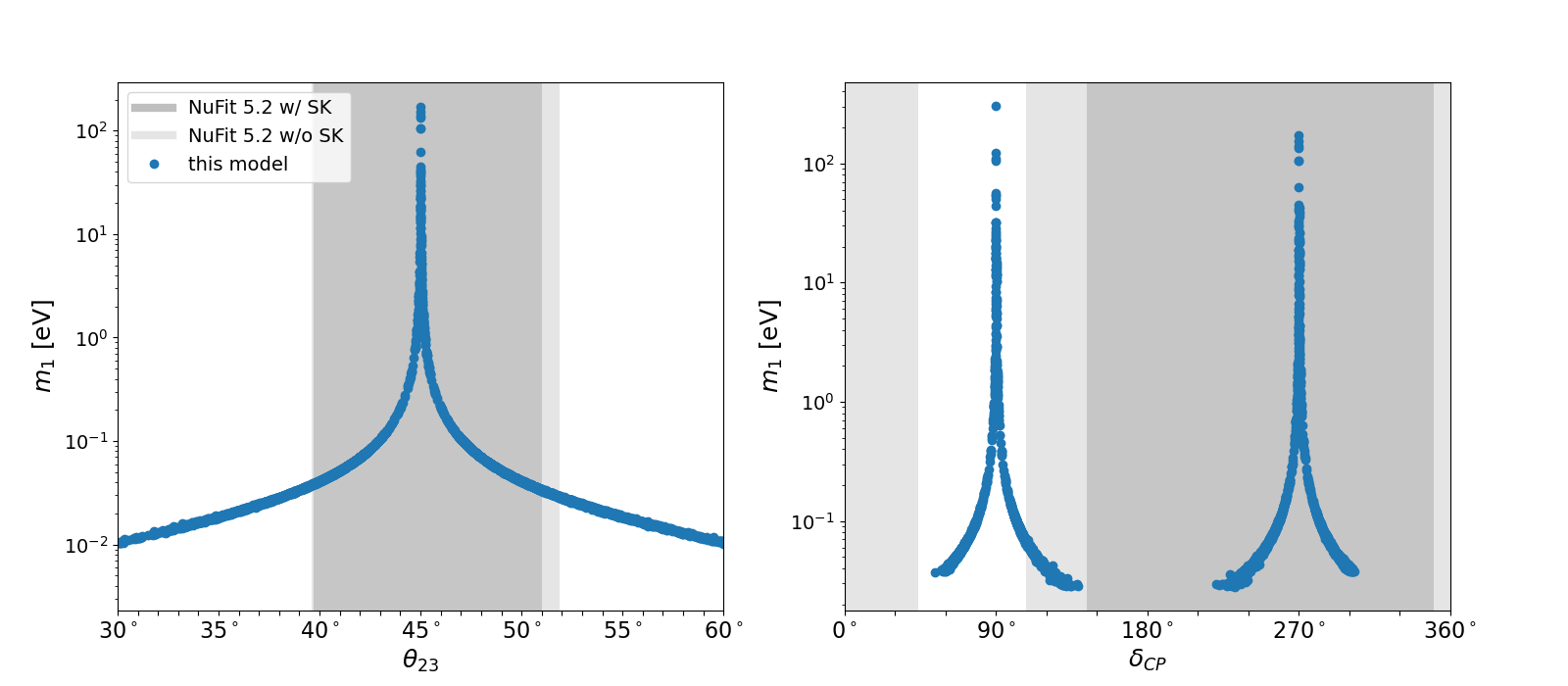}
        \caption{Correlation between the lightest neutrino mass $m_1$ and neutrino mixing parameters $\theta_{23}$ and $\delta_{CP}$. Each data point satisfies that NuFit 5.2 constraints for the remaining neutrino mixing parameters at 3$\sigma$ CL while assuming normal hierarchy.}
    \label{Neutrino_mixing_2}
\end{figure}

An interesting result in the predictions for $\theta_{23}$ and $\delta_{CP}$ is that the atmospheric mixing angle $\theta_{23}$ can attain values near the maximal mixing value $45^\circ$ when $\sum_i m_i$ is sufficiently large. This behaviour is further illustrated in Fig.~\ref{Neutrino_mixing_2}, where the lightest neutrino mass $m_1$ is presented as function $\theta_{23}$ (left panel) and $\delta_{CP}$ (right panel). In both panels, we have excluded the data points that are not consistent with the neutrino oscillation parameter constraints at 3$\sigma$ CL. In this regard, the cuts are performed for the five neutrino oscillation parameters that are not displayed on the horizontal axis. We have again used the experimental constraints that do not account the atmospheric data from SK. The corresponding 3$\sigma$ CL ranges are also shown for $\theta_{23}$ and $\delta_{CP}$ parameters. The results show that $\theta_{23} = 45^\circ$ is achievable in this model when $m_1$ is very large. In a similar manner, maximally violating $CP$ phases $\delta_{CP} = 90^\circ$ and $270^\circ$ are reachable when $m_1 \sim 1$~eV or above. However, both cases are excluded by the cosmological bounds on $\sum_i m_i$.

We finally checked that our results are in agreement with the sum rules that were discussed in Ref.~\cite{Asai:2017ryy}. All of the data points shown in Figures~\ref{Neutrino_mixing_1} and~\ref{Neutrino_mixing_2} are found to be consistent with the sum rules by 3$\sigma$ CL.

\section{Finite temperature}
\label{sec:finiteT}

In this section we introduce the 1-loop effective potential at zero and finite temperature. We then proceed to discuss the appearance of first order phase transitions and the multiple vacuum tunneling patterns in field space. 

In the preceding sections we fixed the gauge sector and neutrino parameters with the $g-2$ and neutrino oscillation data. The remaining free parameters are the dark vector-like fermion mass and charge $m_{\psi}, q_{\psi}$, respectively, and also the parameters in the tree-level scalar potential, Eqn.~\eqref{V0}. For the tree-level scalar potential, we perform a scan with the following uniform ranges:
\begin{equation}
    m_{\phi}/2<m_{\phi'},m_{DM}<500 \ \text{GeV}, \nonumber
\end{equation}
\begin{equation}
    -3<\lambda_{D\phi}, \ \lambda_{D\phi'}<3,  \nonumber
\end{equation}
\begin{equation}
   -0.1 < \theta < 0.1. \nonumber
      \label{angle_range}
\end{equation}
\begin{equation}
    -10^3\leq q_{DM} \leq 10^3.
\end{equation}

In our numerical scans, we verify that the quartic couplings respect the positivity and perturbativity bounds. For the positivity of the potential, we employ the formulae from the appendix of Ref.~\cite{Biswas:2016yan}. We later apply the perturbativity bound $|q_{DM} g_{Z'}|\leq \sqrt{4\pi}$ when we study DM in section~\ref{section_DM}.

Choosing the extra scalar masses above the Higgs threshold allows us to avoid the stringent constraints coming from Higgs invisible constraints. The portal couplings are chosen to be in the natural range, while the mixing angle can easily satisfy the constraints from the Higgs signal strengths measurements. For simplicity, we fix $\lambda_{DM}=1$ in our scans. We have verified that our conclusions remain qualitatively the same for different values of $\lambda_{DM}$. The vector-like fermion properties are found by requiring the correct DM abundance and detection constraints. The corresponding results are also presented in section~\ref{section_DM}.

\subsection{Effective potential }

The zero temperature 1-loop correction is given by the Coleman-Weinberg potential. In the on-shell scheme, the potential can be written as
\begin{equation}
V_{\text{CW}}(\phi,\phi',\chi) = \sum_i (-1)^{F_i}  \frac{d_i}{64 \pi^2} \left[ m_{i}^4(\phi,\phi',\chi) \left(  \log{\frac{m_{i}^2(\phi,\phi',\chi)}{m_{0i}^2}}  - \frac{3}{2}   \right) +2m_i^2(\phi,\phi',\chi) m_{0i}^2 \right] \, , \label{CW}
\end{equation}
where the index $i$ runs over all particles contributing to the potential 
with $F_i=0$ ($1$) for bosons (fermions), $d_i$ is the number of 
degrees of freedom of the particle species, $m_i(h,H,\chi)$ is the 
field-dependent mass of particle $i$ and $m_{0i}$ is the mass at zero temperature, respectively. 

At finite temperature one must add the contribution~\cite{Dolan:1973qd}  
\begin{equation}
V_T(\phi,\phi',\chi, T) = \frac{T^4}{2\pi^2}  \sum_i d_i J_{\mp} \left( \frac{m_i(\phi,\phi',\chi)}{T} \right) \, ,
\label{VT}
\end{equation}
where $d_i$ are the number of degrees of freedom and the $J_{\mp}$ functions are defined as  
\begin{equation}
J_{\mp}(x)  = \pm \int^{\infty}_{0} dy y^2 \log\left( 1 \mp e^{-\sqrt{y^2 + x^2}} \right) \, ,
\end{equation}
and the upper (lower) sign is for bosons (fermions).

The finite temperature effective potential is given by the superposition of Eqns.~\eqref{V0}, \eqref{CW} and \eqref{VT}. It can be written as
\begin{equation}
    V_{eff}(\phi,\phi',\chi,T) =V_0(\phi,\phi',\chi) + V_{CW}(\phi,\phi',\chi) +V_T(\phi,\phi',\chi,T).
\end{equation}

\subsection{Phase transitions}

The non-trivial multi-field potential can easily feature potential barriers that could give rise to temperature fluctuations. These fluctuations can then transition the system from unstable local minima towards the absolute minimum of the free energy. In the theory of vacuum tunneling at finite temperature~\cite{Coleman:1977py,Callan:1977pt,Coleman:1980aw,Linde:1980tt,Linde:1981zj}, the transition probability between such states is given by 
\be\label{Gamma3T}
\Gamma_3(T) =\left(  \frac{S_3(T)}{2\pi T}\right)^{3/2} T^4 e^{-S_3(T)/T}~,
\ee
where $S_3$ denotes the $O(3)$ symmetric Euclidean action of the theory, which are obtained by integrating the equation of motion for the scalars. The phase transition proceeds via bubble nucleation which the associated nucleation temperature $T_n$ reached when the decay rate in Eqn.~\eqref{Gamma3T} becomes comparable to the Hubble volume. This happens when $\Gamma_3 \approx H^4$, where the Hubble scale is written as
\be\label{H2full}
H^2 =  \frac{ \rho_R }{3  M_{\text{Pl}}^2}   + \frac{\Delta V_{eff}(\phi,\phi',\chi,T)}{3 M_{\text{Pl}}^2} ~, 
\ee
with $\rho_R=\pi^2 g_* T^4/30$, the energy density in radiation with $g_*$ number of degrees of freedom \cite{Saikawa:2018rcs}. Furthermore, we take into account the energy density contribution from the scalar potential, which can be especially significant for strong phase transitions where a large amount of latent heat is released. The latent heat released normalized to the radiation energy density is encoded by
\be
\alpha(T) \equiv  \frac{1}{\rho_{R}}\left( \Delta V_{eff}(\phi,\phi',\chi,T) - \frac{T}{4} \Delta \frac{\partial V_{eff}(\phi,\phi',\chi,T)}{\partial T} \right)~.
\ee
In the above formulae, the potential energy difference is between the local and global minima, {\it i.e.}, $\Delta X \equiv X_{\text{local}} - X_{\text{global}} $. The process of bubble nucleation is also characterized by an inverse time scale or inverse time duration $\beta$. This can be written as
\be 
\frac{\beta}{H} \equiv T \frac{d}{dT} \left(  \frac{S_3(T)}{T} \right)\bigg{|}_{T \to T_n}\,.
\label{eq:betaHN}
\ee 
Another crucial property of the nucleation process is the velocity at which the bubble wall interface expands and whether the bubble wall runs away at the speed of light or reaches a steady state velocity. Since we consider a polynomial potential with tree-level mass terms in the present work, the amount of supercooling is very limited and the walls are expected to reach a steady state. Furthermore, since the only particles beyond the SM that can gain or lose mass during the transition are the $U(1)_{L_{\mu} -L_{\tau}}$ gauge boson and the RHN Majorana fermions, the source of friction coming from their distribution functions is also small. We have verified that the changes in the masses of these particles are negligible. In this regard, we take a conservative approach and use the analytic formula~\cite{Lewicki:2021pgr, Ellis:2022lft}
\be 
v_w =
\begin{cases}
\sqrt{\frac{\Delta V}{\alpha \rho_R}} \quad \quad {\rm for} \quad \sqrt{\frac{\Delta V}{\alpha \rho_R}}<v_J(\alpha)~,
\\
1 \quad \quad \quad \quad {\rm for} \quad  \sqrt{\frac{\Delta V}{\alpha \rho_R}} \geq v_J(\alpha) \, ,
\end{cases}
\label{eqn:approx_velocityN}
\ee
where $v_J=\frac{1}{\sqrt{3}}\frac{1+\sqrt{3 \alpha^2+2 \alpha}}{1+\alpha} $ is the Chapman-Jouguet velocity.

\begin{table}[H]
\centering
\begin{tabular}{|c|c|c|c|c|c|c|c|}
\hline
\textbf{direction} & \textbf{All} & \textbf{$\phi-\phi'$} & \textbf{$\phi-\chi$} & \textbf{$\phi'-\chi$} & \textbf{$\phi$} & \textbf{$\phi'$} & \textbf{$\chi$} \\
\hline
\textbf{label} & 0 & 1 & 2 & 3 & 4 & 5 & 6 \\
\hline
\end{tabular}
\caption{All possible patterns of a PT in this model. The \textbf{direction} row indicates which fields underwent changes in their vev. For example, the transitions with \textbf{label} $0$ have all fields changing their vev while those with the label $4$ only the Higgs field vev changes.}
\label{PT_pattern}
\end{table}

As the symmetry of the theory is restored at sufficiently high temperature, the vacuum expectation value of the fields must evolve so that the symmetry is spontaneously broken as the universe cools. Different values of the parameters of the theory can lead to phase transitions with very distinct dynamical evolution of the fields.  We classify all of the possibilities in table~\ref{PT_pattern}.

The results of the parameter scan reveal that a strong first order phase transition can only occur for labels $0$ and $2$. This symmetry breaking pattern corresponds to the condition
\begin{eqnarray}
    \Delta \phi \neq 0 \quad \&  \quad \Delta \chi \neq 0 \quad \& \quad 
 ( \Delta \phi' =0 \quad  || \quad \Delta \phi' \neq 0 ) \, ,
\end{eqnarray}
which means that during the phase transition the change in vacuum expectation value of the $h$ and $\chi$ fields is always different from zero whilst the extra singlet state $\phi'$ may or may not change its vev. Other patterns were also obtained, but they suffer from a very weak transition strength and fall below the crossover regime, $\alpha \approx 10^{-3}$. Below this regime, we expect that a lattice study would reveal a different nature of the transition. Moreover, we find that each PT pattern is in a one-to-one correspondence with a single benchmark point each, thus there are no multi-step PTs.

\begin{figure}
    \centering
\includegraphics[height=0.5\textwidth]{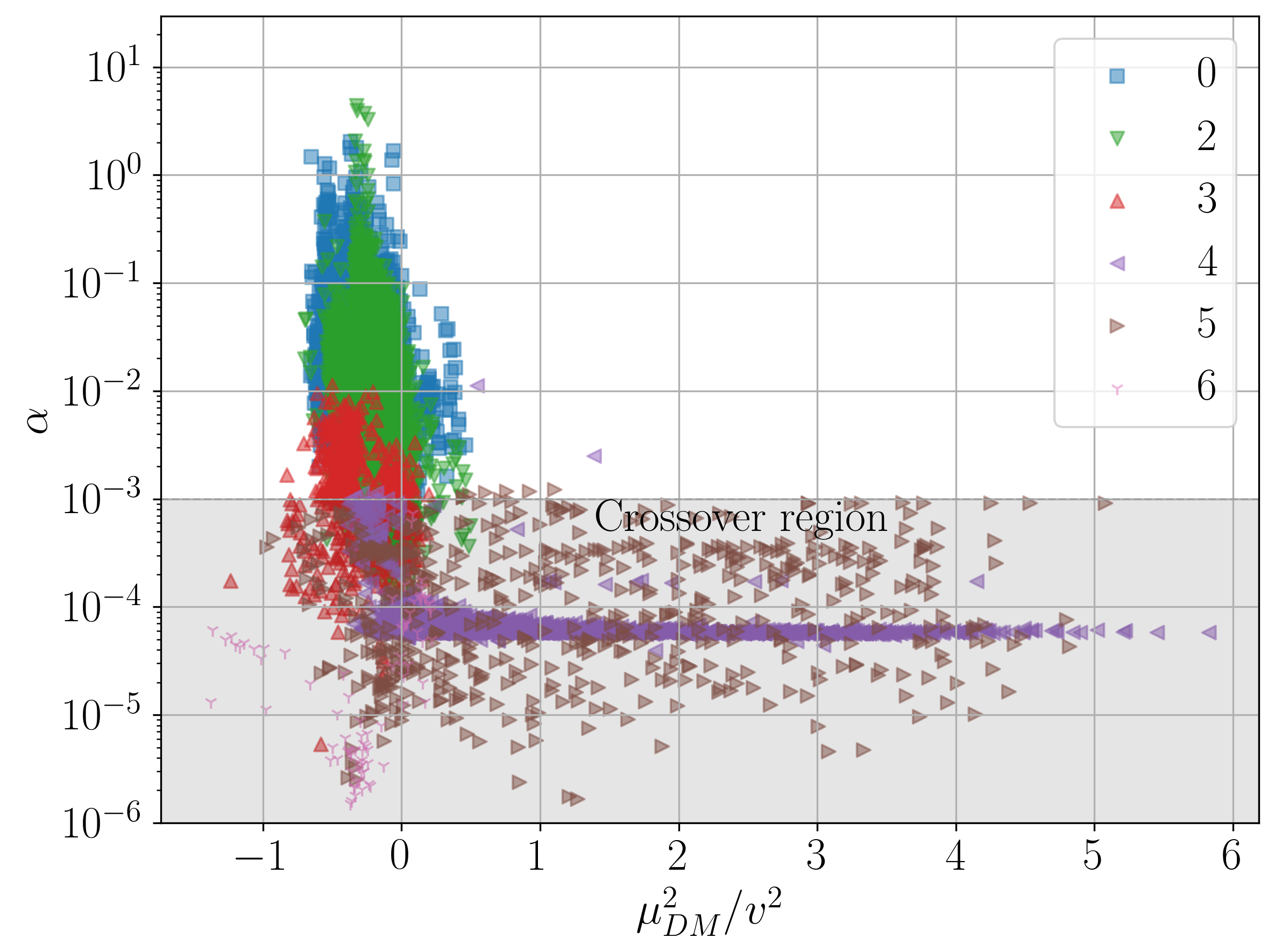}
    \caption{Strength of the PT as a function of DM mass-squared parameter normalized to the EW scale. The color coding of the scatter plot corresponds to the PT patterns of table~\ref{PT_pattern}. The gray area in the figure marks the region with crossover transitions.}
    \label{PT_stength_mass}
\end{figure}

From the data, we inspect the Spearman rank correlation coefficient $\rho$ between the strength and the model parameters. The result is that the DM mass-squared, $\mu_{DM}^2$, has the largest correlation with $\rho = -0.7 $. We display the results in Fig.~\ref{PT_stength_mass}, showing the variation of $\alpha$ with $\mu_{DM}^2/v^2$.

We do not find any PT with label $6$ in the scan, meaning that the transitions happening solely across the DM direction are ruled out. At the same time, the transitions in the Higgs direction only, which are relevant for EW baryogenesis, fall in the crossover region. However, we stress that transitions with labels $0$ and $2$ could also be used in the context of EW baryogenesis, but we leave this topic for another study.

\section{Gravitational waves}
\label{sec:GWs}

\begin{figure}[t]
    \centering
\includegraphics[height=0.5\textwidth]{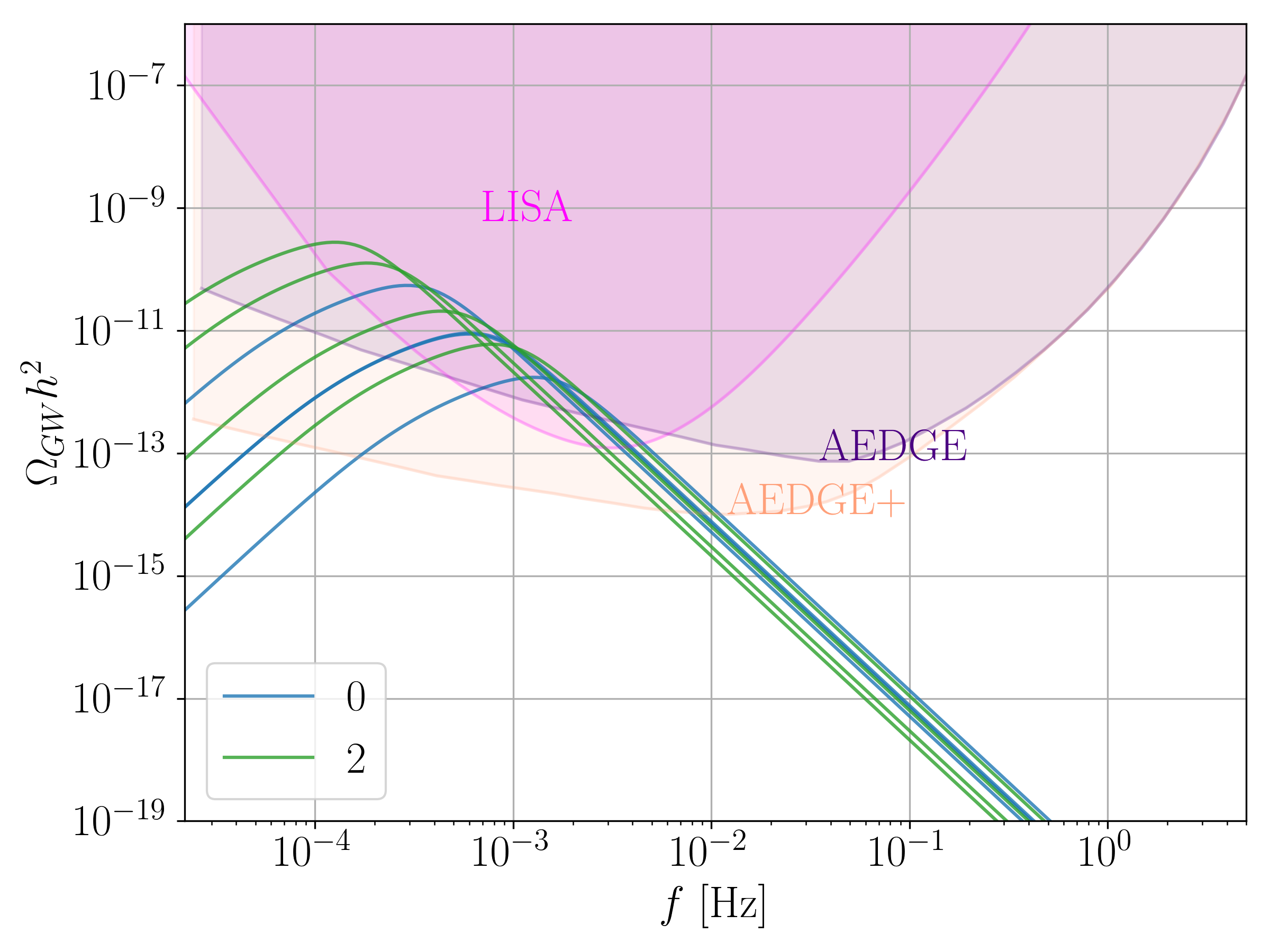}
    \caption{Gravitational wave spectrum for the benchmark points with the strongest phase transitions. The color label of each curve corresponds to each type of transition as displayed in table \ref{PT_pattern}. The prospective experimental sensitivities of various detectors are also shown. }
    \label{GW_signals}
\end{figure}

In this section, we estimate the GW spectrum from bubble dynamics in a first-order phase transition. Several dedicated numerical simulations have been performed in the literature in which the hydrodynamic interactions between the plasma and the accelerating bubble walls are solved on a lattice computation, see Refs.~\cite{Hindmarsh:2013xza,Hindmarsh:2015qta,Hindmarsh:2017gnf,Cutting:2019zws,Hindmarsh:2019phv,Hindmarsh:2016lnk}. The qualitative results of these lattice simulations can be encoded in a template function that describes the GW strain amplitude as a function of the frequency. In the present model, assuming that the potential does not exhibit a significant amount of supercooling, we expect the dominant contribution to be from the sound wave source. We use the sound wave source template from Ref.~\cite{Caprini:2024hue}, which reads
\begin{equation}
    \Omega_{\text{GW}} = \Omega_p  \ S(f),
\end{equation}
where the peak amplitude is obtained as
\begin{equation}
    \Omega_p = 1.64 \times 10^{-5} \left( 100/g_* \right)^{1/3} \left( \frac{0.6 \kappa \alpha}{1+\alpha} \right)^2 (8 \pi)^{1/3}\text{max}(v_w,c_s) \left( \frac{\beta}{H_*}\right)^{-1}\text{min}(H_* \tau_{sh},1).
\end{equation}
Here the subscript $*$ indicates that quantities are evaluated at the production time. In our case, this corresponds to the temperature at which the bubbles of the new phase percolate $T_p \approx T_n$. For the sound speed in the plasma, we take $c_s^2 =1/3$. The efficiency factor $\kappa$ is taken from the seminal work \cite{Espinosa:2010hh}. The estimated decay time into turbulence is encoded in the variable $H_*\tau_{sh} =\sqrt{\frac{4}{3}\frac{1+\alpha}{0.6 \alpha}} (8\pi)^{1/3}\text{max}(v_w,c_s) \left( \beta/H_*\right)^{-1} $.

The spectral function adopts the form of a double-broken power law 
\begin{equation}
    S(f) = N \left( \frac{f}{f_1} \right)^3  \left(1+\left( \frac{f}{f_1} \right)^2 \right)^{-1} \left(1+ \left(\frac{f}{f_2} \right)^4 \right)^{-1},
\end{equation}
where the normalization factor is found by requiring $\int_{-\infty}^{\infty}d\log f S(f)=1$, 
while the frequency breaks are obtained as 
\begin{equation}
 f_1 = 1.13 \times 10^{-6} \left(\frac{g_*}{100} \right)^{1/6} \left(\frac{T_*}{100 \text{GeV}} \right) \left( \frac{\beta}{H_*} \right) \frac{1}{\text{max}(v_w,c_s)},  \quad   \frac{f_2}{f_1} = \frac{5}{2}  \frac{\text{max}(v_w,c_s)}{|v_w-c_s|}.
\end{equation}

We display the associated GW signals in Fig.~\ref{GW_signals} and we superimpose the projected sensitivities of LISA and AEDGE \cite{AEDGE:2019nxb}. We only display the predictions coming from benchmarks with very strong transitions strengths. These correspond to transition patterns of labels $0$ and $2$. 

\section{Dark matter}
\label{section_DM}

As we have discussed before, the $U(1)_{L_{\mu}-L_{\tau}}$ gauge symmetry takes an important role
in explaining the neutrino oscillation parameters and the muon ($g-2$) by its implications on the structure of the neutrino mass and by the presence of the extra gauge boson. In this section, we examine another important effect in determining the DM relic density. Before discussing this further, we first comment on the need for the two-component DM. In principle, one can have a single-component scalar DM and we can solve DM relic density bound from the Planck data. However, as we have shown in the GW section above, a strong first-order PT would require high values of $\lambda_{D\phi}$ and $\lambda_{D\phi^{\prime}}$,
which are already in tension with the direct detection data obtained by the LUX-ZEPLIN collaboration \cite{LZ:2022lsv}. One method to overcome this potential constraint from the direct detection of scalar DM is to introduce a second DM candidate, a dark vector-like fermion $\psi$. The fermionic DM component would be present in the Universe in higher amounts and reduce the effective direct detection cross-section for the scalar DM by a multiplicative factor $\frac{\Omega_{\Phi_{DM}}}{\Omega_{\Phi_{DM}} + \Omega_{\psi}}$. The detailed description of the model has been addressed in the model, see section~\ref{sec:model}.
To study the two-component DM scenario, we have implemented our model file in \texttt{FeynRules}~\cite{Alloul:2013bka} and then generated the \texttt{CalcHEP}~\cite{Belyaev:2012qa} files to use in \texttt{micrOMEGAs}~\cite{Belanger:2001fz}. Moreover, we compare our output from \texttt{micrOMEGAs} with our analytical estimates. In this section, we first discuss the analytical estimate and then proceed to present our results, where the correlations are shown among the different model parameters generated using \texttt{micrOMEGAs}. 

The abundance of DM can be obtained after solving the following Boltzmann equation,
\begin{eqnarray}
    \frac{d Y_{A}}{d x} &= & - \frac{S(x)\,\, \langle \sigma v \rangle}{x H(x)} 
    \left( Y^2_{A} - Y^2_{A, eq} \right)\,,
    \label{eq:Boltzmann_equation}
\end{eqnarray}
where $A = \psi, \Phi_{DM}$, $x = \frac{m_{\psi,DM}}{T}$, and $T$ is the temperature of the Universe. The entropy ($S(x)$) and the Hubble parameter ($H(x)$) can be expressed as
\begin{eqnarray}
S(x) = \frac{2 \pi^{2}}{45} g_{s} m^3_{A} x^{-3}\,,
  H(x) = \frac{\pi}{3} \left( \frac{g_{\rho}}{10} \right)^{1/2} \frac{m^2_{A}}{M_{Pl}} x^{-2}\,,    
\end{eqnarray}
where $m_{A}= m_{\psi}, m_{DM}$ is the DM mass, $g_{s}$ and $g_{\rho}$ are the relativistic d.o.f associated with the entropy and the matter. Although $g_{s}$ and $g_{\rho}$ slightly vary during the evolution of the Universe, we consider them to be equal to the effective d.o.f. defined as $\sqrt{g_{*}} = \frac{g_{s}}{\sqrt{g_{\rho}}} 
\left( 1 + \frac{1}{3} \frac{T}{g_{s}} \frac{d g_{s}}{d T} \right)$~\cite{Gondolo:1990dk}. 
In the above equation, we can solve in a two-step process by determining the freeze-out temperature 
$T_{f} = \frac{m_{A}}{x_{f}}$. The freeze-out temperature represents the temperature 
of the Universe during its evolution when the DM interaction strength falls below the
Hubble expansion rate which can be determined as
\begin{eqnarray}
    n_{A}\, \langle \sigma v \rangle_{AA} = H \,\,\,{\rm at} \,\,\, x = x_{f}\,,
\end{eqnarray}
where the number density of DM $A$  is $n_{A} = \frac{g_{A}}{\pi^{2}} m^3_{A} x^{-3/2} e^{-x}$, $g_{A}$ is internal d.o.f of particle $A$, and the effective cross-section times velocity in the non-relativistic limit can be expressed as $\langle \sigma v \rangle = a_{A} + \frac{6 b_{A}}{x} $\, where $a_{A}$ and $b_{A}$ are the $s-$wave and $p-$wave annihilation modes. After applying the aforementioned expression, we obtain the freeze-out temperature by solving the following equation,
\begin{eqnarray}
    \label{eq_DM1}
    x = 23.64 + \frac{1}{2} {\rm ln}\,{x} + {\rm ln} \left( \frac{m_{A}}{100} \right)
    + {\rm ln}\, g_{A} 
    + \frac{1}{2} {\rm ln} \left[ \frac{\langle \sigma v \rangle}{2.6 \times 10^{-9}\,\,{\rm GeV^{2}}} \right]
    - \frac{1}{2}\,{\rm ln} \left[ \frac{g_{*}}{106} \right].
\end{eqnarray}

After solving Eqn.~\eqref{eq_DM1}, we can determine the freeze-out temperature and thereby solve the Boltzmann
equation as given in Eqn.~\eqref{eq:Boltzmann_equation}. By calculating the evolution of the Boltzmann equation 
from $x = x_{f}$ to $x = \infty$, we can estimate the amount of the co-moving number
density as
\begin{eqnarray}
    Y_{A} = \frac{Y_{A,\,eq} (x_{f})}{1 + Y_{A,\,eq} \left[ \frac{2 \pi}{15} \sqrt{\frac{10}{g_{\rho}}}
    \, g_{s} m_{A} M_{\text{Pl}} \right] \,\frac{1}{x_{f}} \left( a_{A} + \frac{6b_{A}}{x_{f}} \right) \kappa      },
    \label{eq:comoving_density}
\end{eqnarray}
where $\kappa = 1/2$ in case the particle and anti-particle are not the same. Otherwise $\kappa = 1$~\cite{Gondolo:1990dk}. 

The factors $a_{A}$ and $b_{A}$ can be calculated for the $\Phi_{DM}$ and $\psi$ DM candidates. We present here
only the DM annihilation to the extra gauge bosons, which is the dominant mode in the present work. The result
reads as follows,
\begin{eqnarray}
    a_{\Phi_{DM}} &\simeq& \frac{{g^4_{Z'} } {q^4_{DM}}}{\pi  \left(8 {m^2_{{DM}}}\right)}-\frac{{g^4_{Z'}} {m^2_{Z'}} {q^4_{DM}}}{\pi  \left(16 {m^4_{{DM}}}\right)}\,, \nonumber \\
        b_{\Phi_{DM}} &\simeq& -\frac{11{g^4_{Z'}  } {q^4_{DM}}}{\pi  \left(192 {m^2_{{DM}}}\right)}+\frac{3{g^4_{Z'}} {m^2_{Z' }} {q^4_{DM}}}{\pi  \left(128 { m^4_{{DM}} }\right)}\,, \nonumber \\
     a_{\psi} &\simeq& \frac{g^{4}_{Z'} q^{4}_{\psi}}{16 m^2_{\psi} \pi } 
     - \frac{g^{4}_{Z'} q^4_{\psi} m^2_{Z' } }{32 m^{4}_{\psi} \pi }
     - \frac{3 g^{4}_{Z'} q^4_{\psi} m^4_{Z' } }{128 m^{6}_{\psi} \pi }\,, \nonumber \\
     b_{\psi} &\simeq& \frac{3 g^{4}_{Z'} q^{4}_{\psi}}{128 m^2_{\psi} \pi } 
     + \frac{29 g^{4}_{Z'} q^4_{\psi} m^2_{Z' } }{768 m^{4}_{\psi} \pi }
     + \frac{23 g^{4}_{Z'} q^4_{\psi} m^4_{Z' } }{1024 m^{6}_{\psi} \pi }\,.
     \label{a-b-expression}
\end{eqnarray}
Once we obtain the co-moving number density $Y_{A}$, we can determine the total DM density,
\begin{eqnarray}
    \Omega_{DM}h^{2} = \sum_{A = \Phi_{DM}, \psi} \Omega_{A} h^{2} = \sum_{A = \Phi_{DM}, \psi} 2.755 \times 10^{8}\, 
    Y_{A}\,\left( \frac{m_{A}}{\rm GeV} \right)\,.
    \label{eq:total_relic_density}
\end{eqnarray}

We have checked that the analytical result obtained using the above expressions and 
\texttt{micrOMEGAS} match each other with less than $10\,\%$ discrepancy\footnote{The discrepancy further decreases once
we multiply $\langle \sigma v \rangle$ by $\frac{1}{2} 
\left( 1 + \frac{K_{1}(x)}{K_{2}(x)} \right)$, where $K_{i}(x)$ is the modified Bessel function of the second kind~\cite{Gondolo:1990dk}.}. In our study, we have used \texttt{micrOMEGAS} to generate the DM relic density and its associated
observables. In this respect, we have varied the model parameters within the following ranges,
\begin{eqnarray}
  && \frac{m_{\phi}}{2} \leq m_{\phi'} \ [\text{GeV}] \leq 2000\,,\,\,
  10^{-4} \leq m_{Z' } \ [\text{GeV}] \leq 1\,,\,\,
  1 \leq m_{\psi} \ [\text{GeV}] \leq 1000\,,\,\, \nonumber \\
  && 10^{-6} \leq g_{Z'}  \leq 10^{-1}\,,\,\,
  10^{-4} \leq \sin\theta \leq 0.5\,,\,\,
  1 \leq q_{DM} \leq 10^{5}\,,\,\, \nonumber \\
  && 1 \leq q_{\psi} \leq 10^{5}\,,\,\,
  10^{-2} \leq \lambda_{D\phi} \leq 3\,,\,\,
  10^{-2} \leq \lambda_{D\phi'} \leq 3\,, -2 \leq \frac{\mu^2_{DM}}{v^{2}} < 4 \,.
\end{eqnarray}
It is worth mentioning that we have chosen the $g_{Z'}$ and $m_{Z'}$ values to explain the muon $(g-2)$, which requires small $g_{Z'}$ values, {\it i.e.}, $g_{Z'} \sim 10^{-3}$. Therefore, in order to have a dominant contribution of scalar and fermion DM to the gauge boson, we have considered relatively larger values for 
their $U(1)_{L_{\mu} - L_{\tau}}$ charges, $q_{DM}$ and $q_{\psi}$.

\subsection{Dark matter bounds}
\label{dm-bounds}
Below are listed the points that satisfy the bounds associated with the DM.
\begin{itemize}
    \item {\bf DM relic density bound:}
In the present work, we have considered the Planck bound on the total DM relic density coming from the fermion and scalar sectors. In particular, we have considered $7\sigma$ variation in DM relic density as shown below \cite{Planck:2015fie,Planck:2018vyg},
\begin{eqnarray}
    \Omega_{DM}h^{2} = \left( \Omega_{\psi} + \Omega_{\Phi_{DM}} \right) h^{2} = 0.1200 \pm 0.0084\,.
\end{eqnarray}
We consider a wide range for DM relic density around its central value $\Omega_{DM}h^{2} = 0.12$ to get a large number of points in reasonable time while satisfying the DM relic density. Furthermore, if we consider the narrower $1\sigma$ range in DM relic density as given in the Planck paper~\cite{Planck:2015fie,Planck:2018vyg} instead, we would not expect any change in our conclusion. In such case, a slight reduction in the allowed parameter space in different planes would be expected. The main diagrams determining the relic densities of $\phi_{DM}$ and $\psi$ DM candidates are shown in Fig.~\ref{dm-annihilation}.

\begin{figure}[t!]
\centering
\includegraphics[scale=.65]{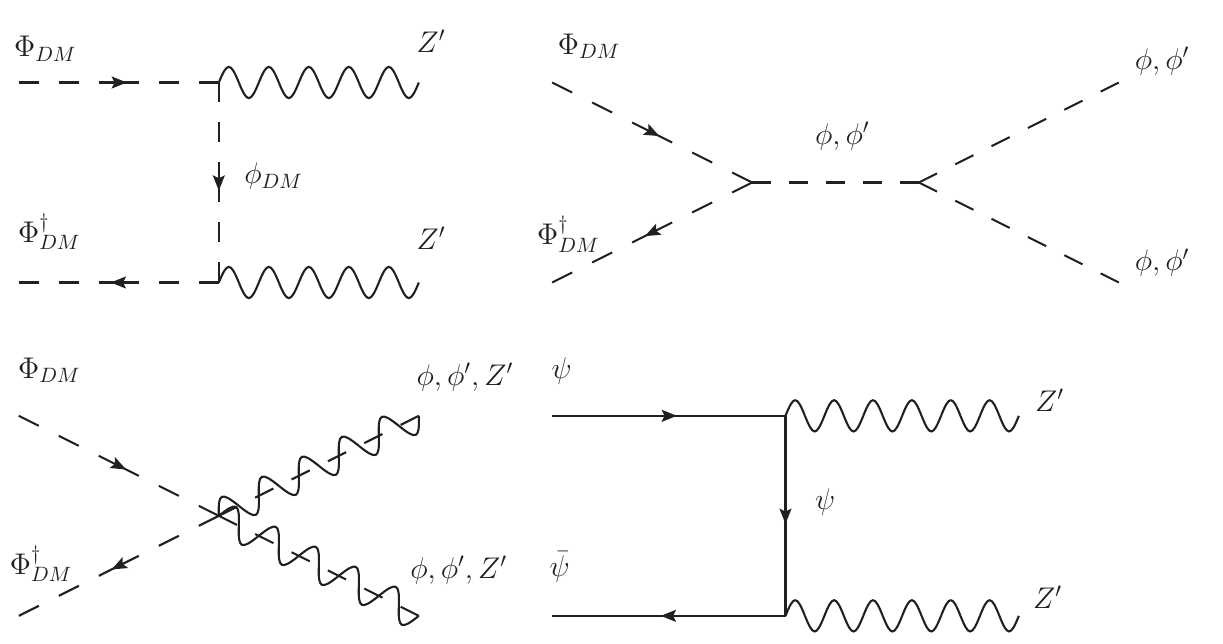}
\caption{Main annihilating diagrams for the DM candidates, $\Phi_{DM}$ and $\psi$, that determine
their relic density.}
\label{dm-annihilation}
\end{figure}

\item {\bf DM Direct detection:} In this work, we have considered direct detection bound on the scalar DM and no bound applied to the fermion DM due to its inert nature. Scalar DM can interact with the nucleon of the detector as shown in Fig.~\ref{dd-id}. The spin-independent direct detection (SIDD) cross-section for the scalar DM can be estimated in the non-relativistic limit as
\begin{eqnarray}
    \sigma_{SI} = \frac{\mu^2_{red} f^2_{N} M^2_{N} }{ 4 \pi m^2_{DM} v^{2}} 
    \left( \frac{g_{\Phi^{\dagger}_{DM} \Phi_{DM} \phi}\, \cos\theta }{m^2_{\phi}}
    - \frac{g_{\Phi^{\dagger}_{DM} \Phi_{DM} \phi' }\, \sin\theta }{m^2_{\phi'}} \right)^{2}\,,
    \label{si-dd-cs}
\end{eqnarray}
where $\mu_{red} = \frac{m_{DM} M_{N}}{m_{DM} + M_{N}}$ is the reduced mass between DM and nucleon, 
$M_{N}$ is the nucleon mass, $f_{N} \sim 0.3$ \cite{Junnarkar:2013ac} 
is the nucleon form factor and the DM couplings 
with the Higgs bosons taking the following form,
\begin{eqnarray}
    g_{\Phi^{\dagger}_{DM} \Phi_{DM} \phi' } &=&  \left[ \lambda_{D\phi' } v' \cos\theta 
    + \lambda_{D\phi} v \sin\theta \right], \,\nonumber \\
    g_{\Phi^{\dagger}_{DM} \Phi_{DM} \phi } &=&  \left[ \lambda_{D\phi' } v' \sin\theta 
    - \lambda_{D\phi} v \cos\theta \right] .
    \label{dm-dm-phi-phiP}
\end{eqnarray}
In the present work, we have considered the most recent bound coming from the measurement by the LUX-ZEPLIN collaboration~\cite{LZ:2022lsv}. All the points which will be shown in the later plots are obtained after satisfying the LUX-ZEPLIN data. Since we have a component DM scenario, one of the components can be small and the other component can be large. Therefore, while comparing with the LUX-ZEPLIN data we have used an effective SIDD cross-section $f_{\Phi_{DM}} \sigma_{SI}$, where $f_{\Phi_{DM}}$ is the fraction of scalar DM present in the Universe.

\item {\bf DM Indirect detection:}
In the present work, the scalar DM can also annihilate to the SM particle through Higgses mediator process as $\Phi^{\dagger}_{DM} \Phi_{DM} \rightarrow \bar f f, VV$ where f is SM fermions and $V$ is the SM gauge bosons. One can estimate the indirect detection cross-section by using the following relation,
\begin{eqnarray}
    \langle \sigma v \rangle_{\Phi_{DM}} = \frac{1}{8 m^4_{{DM}} T K^2_{2} \left( \frac{m_{{DM}}}{T} \right) } \int_{4 m^2_{{DM}}}^{\infty} \sigma_{\Phi^{\dagger}_{DM} \Phi_{DM} \rightarrow f \bar f } 
\left( s - 4 m^2_{{DM}} \right) \sqrt{s} K_{1} \left( \frac{\sqrt{s}}{T} \right)\,ds\,. \nonumber \\
\end{eqnarray}
In the present work, DM mainly annihilates to the extra gauge boson $Z^{\prime}$ and the Higgses $\phi, \phi'$. Hence, DM annihilation to the SM visible sector is suppressed. It also does not effectively take part in the DM relic density calculation. Therefore, it is hard to detect DM at the present with the indirect detection techniques because of the suppressed direct annihilation. Although there will be a possibility to detect DM by indirect detection from the cascade decays 
of the final states gauge boson and Higgses, which we expect to be suppressed. The diagrams associated with the indirect detection of $\phi_{DM}$ and $\psi$ 
are shown in Fig.~\ref{dd-id}.

\begin{figure}[t!]
\centering
\includegraphics[scale=.65]{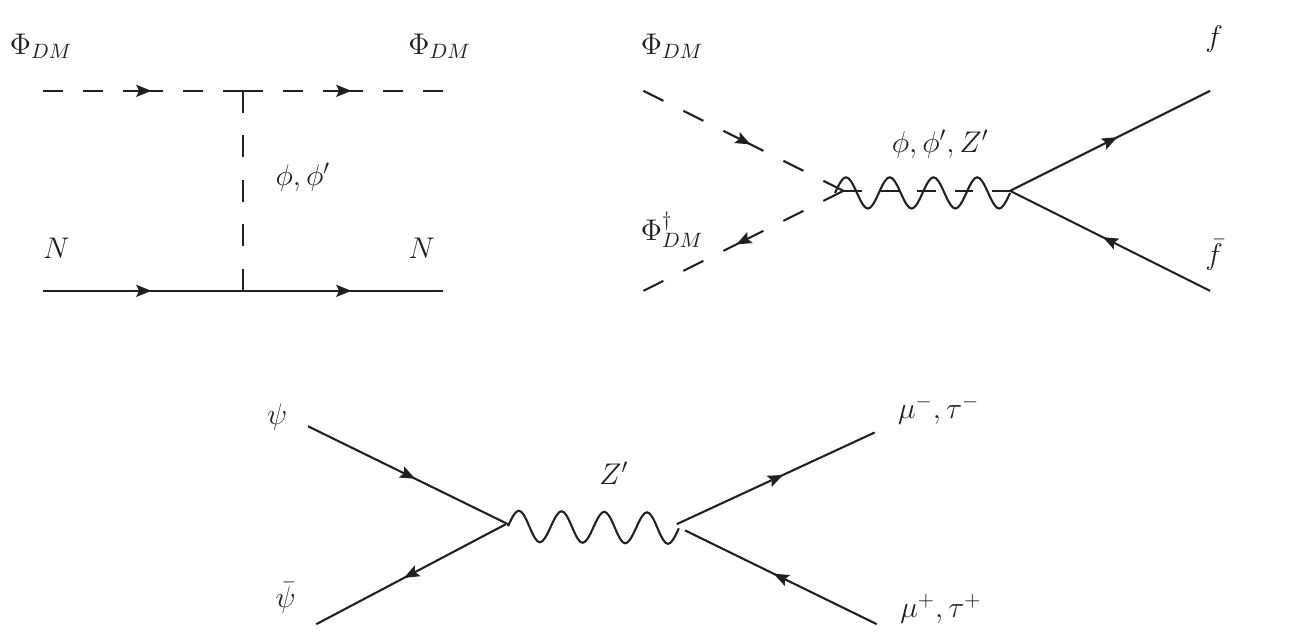}
\caption{DM direct detection and indirect detection for the scalar and fermionic 
DM candidates. }
\label{dd-id}
\end{figure}

\item {\bf Collider bounds:}

We have considered the relevant bound which may appear from the different collider searches. We have taken into account the relevant bounds that can appear for the SM and BSM Higgs by using the \texttt{HiggsSignal}~\cite{Bechtle:2013xfa} 
and \texttt{HiggsBounds}~\cite{Bechtle:2015pma} 

packages. The experimental constraints that are important for the present work are Higgs signal strength, its decay to the invisible sector and BSM Higgs searches at the collider through its decay to the visible sector. We have used the in-built \texttt{HiggsBound} and \texttt{HiggsSignal} packages in \texttt{micrOMEGAs} by externally feeding with the appropriate model file with the relevant couplings.
\end{itemize}

\subsection{Allowed parameter regions from dark matter}

We are now are ready to present the scatter plots that display the correlation between relevant parameters. We have ensured that the results satisfy all of the experimental constraints described in section~\ref{dm-bounds}. We obtain a very nice correlation between the parameters and the observables, as we will describe below in detail. The main contributing diagrams which determine the $\phi_{DM}$ and $\psi$ DM relic densities are shown in Fig.~\ref{dm-annihilation}.

\begin{figure}[t!]
\centering
\includegraphics[angle=0,height=7.5cm,width=7.5cm]{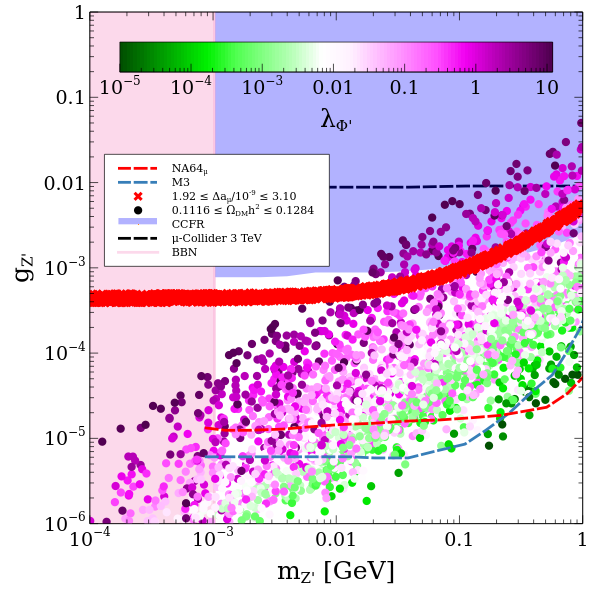}
\includegraphics[angle=0,height=7.5cm,width=7.5cm]{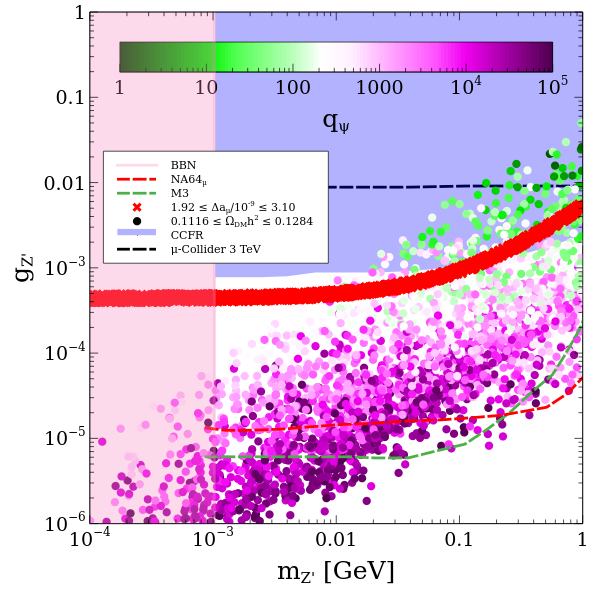}\\
\includegraphics[angle=0,height=7.5cm,width=7.5cm]{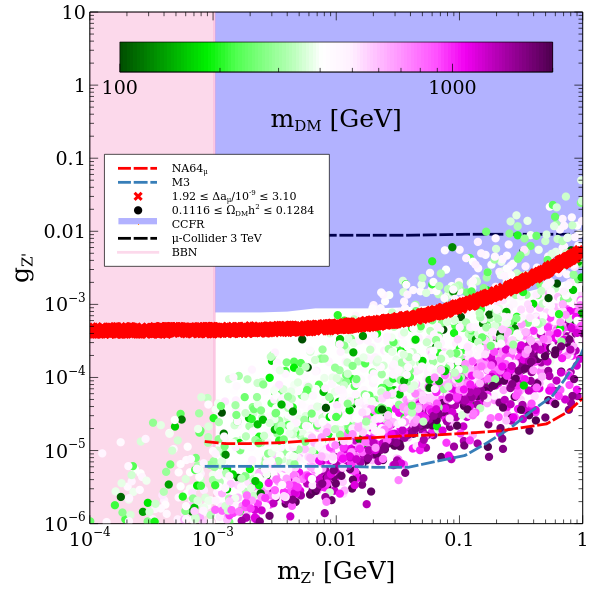}
\includegraphics[angle=0,height=7.5cm,width=7.5cm]{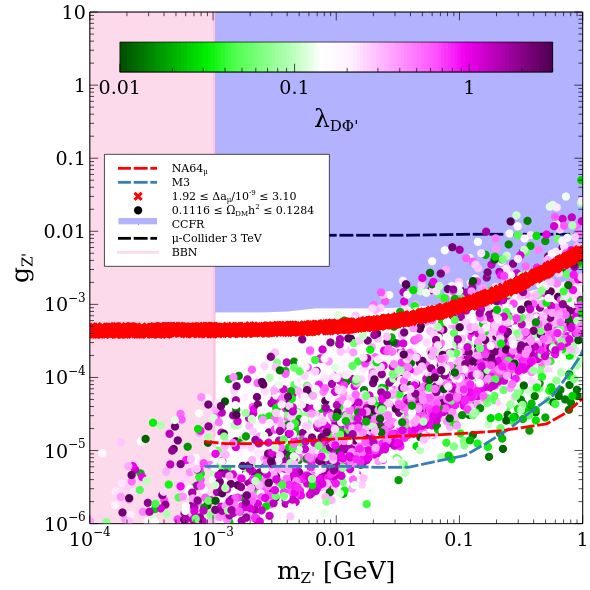}
\caption{Scatter plots in the $m_{Z^{\prime}} - g_{Z'}$ plane after satisfying the DM relic density bound. The colorbars in the four plots represent quartic coupling $\lambda_{\phi'}$, charge of the extra fermion $q_{DM}$, scalar DM mass $m_{DM}$ and quartic coupling $\lambda_{D\phi'}$. We have shown the bounds coming from the neutrino trident experiment CCFR, BBN and future access by the muon collider, M3 and NA64$_{\mu}$.}
\label{DM-plot-1}
\end{figure}

In Fig.~\ref{DM-plot-1}, we show the scatter plots for the $m_{Z^{\prime}}-g_{Z^{\prime}}$ plane where the colour variation corresponds to the different values of $\lambda_{\phi'}$, $q_{\psi}$, $m_{DM}$ and $\lambda_{D\phi'}$ from the top left to the bottom right figures, respectively. While generating the plots, we have checked that all constraints relevant for DM are satisfied as discussed in Section~\ref{dm-bounds}. In the top left panel, we show the colour variation in the quartic coupling $\lambda_{\phi'}$. As can be seen in Eq.~\eqref{quartic-couplings}, the quartic coupling varies with the vev of singlet scalar as $\lambda_{\phi'} \propto \frac{1}{v'^{2}} = \frac{g^2_{Z^{\prime}}}{m^{2}_{Z^{\prime}}}$. One can clearly see the colour variation follows the aforementioned equation. For example, if one fixes $g_{Z^{\prime}}$ and increases the value of $m_{Z^{\prime}}$, then a decrement in the $\lambda_{\phi'}$ appears. Similarly, if one fixed $m_{Z^{\prime}}$ and increased the value of $g_{Z^{\prime}}$ in increments, then one would see an increment in the quartic coupling. There are no points in this panel that would involve high values of $g_{Z^{\prime}}$ and low values of $m_{Z^{\prime}}$ due to the pertubativity bound on the quartic coupling $\lambda_{\phi'} < 4 \pi$, as can be seen above the magenta points. Furthermore, we also have no points below the green points, which can be better understood by looking at the lower panel. In the figure, red patch explains the muon ($g-2$) anomaly. We can see that $m_{Z^{\prime}} > 0.1$ GeV is already in conflict with the neutrino trident bound obtained from CCFR experiment~\cite{Altmannshofer:2014pba}, which is depicted with the blue region. The light brick region shows the values that are ruled out by BBN~\cite{Ahlgren:2013wba}, whereas the dashed black line, dashed red line and dashed light blue line represent the future sensitivities of the muon collider~\cite{MuonCollider:2022xlm}, NA64$_{\mu}$~\cite{Gninenko:2014pea} and M3 experiments~\cite{Kahn:2018cqs}, respectively. In the upper right panel, we show the same scatter plot but with the colour variation corresponding to the $U(1)_{L_{\mu} - L_{\tau}}$ charge $q_{\psi}$. It can be seen that for $q_{\psi} \sim 100$, the model can satisfy both the DM and muon $(g-2)$ constraints at the same time, whilst the smaller values of $q_{\psi}$ are ruled out by the CCFR bound and the larger values correspond to the the smaller contribution to the $(g-2)_{\mu}$ anomaly. In the bottom left and bottom right panels, it can be understood that the absence of points below the magenta points in the bottom left panel and green points in the bottom right panel. This mainly results from the presence of the upper bound on the scalar DM, which we have considered in the present work to be $m_{DM} < 2000$ GeV. Therefore, the aforementioned effect comes from our choice of the model parameters. As we have shown before, $m^2_{DM} \propto \lambda_{D\phi' } v'^2$, meaning that for the values of $v'$ obtained in those regions one needs $\lambda_{D\phi' } < 0.01$. This is the reason why we did not obtain any points in the lower right corners in the panels. This can be seen in the bottom right panel in particular, where green points corresponding to $\lambda_{D\phi' } > 0.01$ can be found.

\begin{figure}[t!]
\centering
\includegraphics[angle=0,height=7.5cm,width=7.5cm]{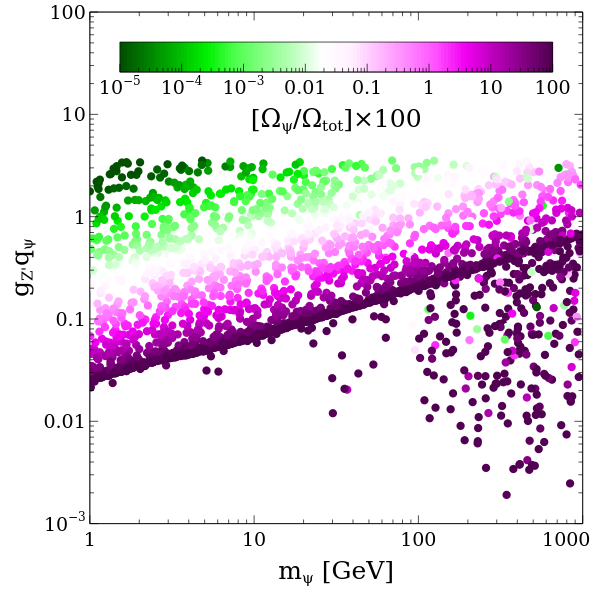}
\includegraphics[angle=0,height=7.5cm,width=7.5cm]{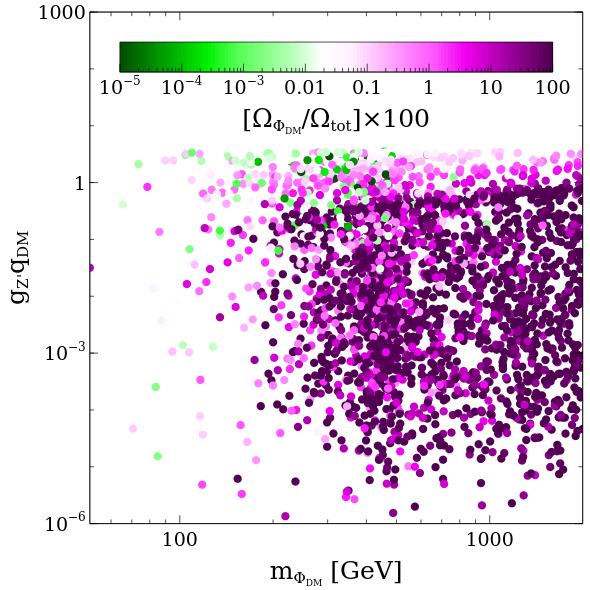}
\caption{Left (light) panel exhibits the scatter plot in the $m_{\psi} - g^{\prime} q_{\psi}\,\,(m_{DM} - g^{\prime} q_{DM})$ plane. In the left panel (right panel), the colorbar shows the amount of fermionic (scalar) DM contribution compared to the total amount of DM in percentage.} 
\label{DM-plot-2}
\end{figure}

In the left panel of Fig.~\ref{DM-plot-2}, we have shown the scatter plot in the $m_{\psi}-g_{Z^{\prime}} q_{\psi}$ plane. Eqns.~\eqref{eq:comoving_density}, \eqref{a-b-expression} and \eqref{eq:total_relic_density} let us infer that DM relic density varies as $\Omega_{\psi} h^{2} \propto \frac{m_{\psi}}{(q_{\psi} g_{Z^{\prime}})^{4}}$. To get the correct value of DM relic density, one needs to look at the sharp magenta line which gives
$100 \%$ contribution to the DM density from $\psi$. If one looks at the higher values of $q_{\psi} g_{Z^{\prime}}$, there is a lower contribution from $\psi$ in the total DM contribution. The upper bound on the $q_{\psi} g_{Z^{\prime}}$ comes from the perturbativity bound, which is given as $q_{\psi} g_{Z^{\prime}} < \sqrt{4 \pi}$. In the right panel, if we take a fixed value of $q_{\psi} g_{Z^{\prime}}$ and go towards the right end for higher values DM mass $m_{\psi}$, we can then see a higher contribution from from $\psi$ DM as depicted by the magenta points. These points are in agreement with the analytical expression shown before. So far the behaviour of DM is mainly controlled by the $\bar \psi \psi \rightarrow Z^{\prime} Z^{\prime} $. As one crosses the $m_{\psi} > 60 $ GeV region, however, one then has also lower values of $g_{Z'}^{\prime} q_{\psi}$. This is due to the fact that for those points one has a new annihilation mode of DM. In other words, $\bar \psi \psi \rightarrow \Phi^{\dagger}_{DM} \Phi_{DM}$ opens up and has the dominant contribution. The aforementioned points are also affected by the $q_{DM}$ values. In the right panel, we have shown a similar plot for the scalar DM candidate. We can thus draw the same conclusion for the present figure as well, but we can also see that very lower values of $g_{Z^{\prime}} q_{DM}$ are allowed, which was not possible for the left panel. This is because those lower values do not contribute to DM relic density significantly instead $\Phi^{\dagger}_{DM} \Phi_{DM} \rightarrow \phi\phi, \phi' \phi' , \phi \phi', f\bar{f}$ annihilation modes are active which are mainly governed by the $\lambda_{D\phi}, \lambda_{D \phi'}$ quartic couplings.

\begin{figure}[t!]
\centering
\includegraphics[angle=0,height=7.5cm,width=7.5cm]{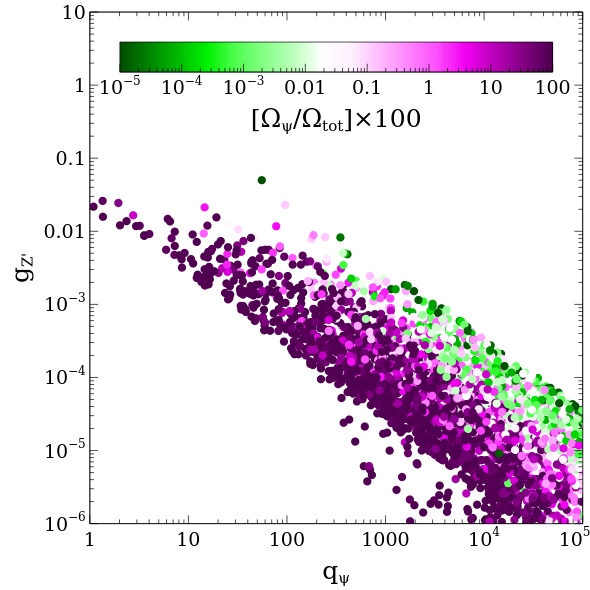}
\includegraphics[angle=0,height=7.5cm,width=7.5cm]{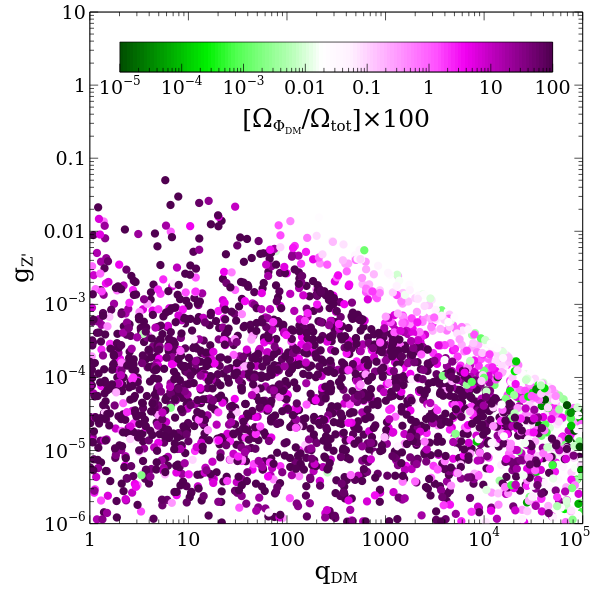}
\caption{Left and right panels show the scatter plots in the $q_{\psi} - g^{\prime}$ and $q_{DM} - g^{\prime}$ planes, respectively. In the left (right) panel, the color variation shows the amount of fermionic (scalar) DM contribution with respect to the total DM density in percentage.} 
\label{DM-plot-3}
\end{figure}
In the left panel and right panel of Fig.~\ref{DM-plot-3}, we show the scatter plots in the $q_{\psi} - g_{Z^{\prime}}$ and $q_{DM}-g_{Z^{\prime}}$ planes, respectively. For a better understanding of the parameter dependence on the DM relic density, we also show the fraction of the fermion DM present as percentage of the total DM contribution in the left panel and for the scalar DM in the right panel. In the left panel, we can see that $q_{\psi}$ and $g_{Z^{\prime}}$ are anti-correlated, which means that for a particular value of $q_{\psi} g_{Z^{\prime}}$, we get $\psi$ contribution in the total DM relic density which does not evade the Planck relic density bound. In the left panel, we can see for a fixed value of $q_{\psi}$, increasing the value of $g_{Z^{\prime}}$ would lead to a lower contribution of $\psi$ DM in the total DM density. This can be clearly seen from the colour variation in the left panel. This behaviour can be easily understood from the analytical estimate, which we have presented in Eq.~\eqref{a-b-expression}. On the other hand, the region above the green points is ruled out as a result of the perturbativity bound on $g_{Z^{\prime}} q_{\psi}$, whereas the region below the deep magenta points is ruled out due to overproduction of $\psi$ DM. In the right panel, we have shown a scatter plot in the
$q_{DM}-g_{Z^{\prime}}$ plane and colour variation shows the fraction of scalar DM in percentage. Again, we find the region above the green points to be ruled out because of the perturbativity bound $g_{Z^{\prime}} q_{DM} < \sqrt{4 \pi}$, but we do not have any constraints from below. This happens since the lower region is dominated by the DM annihilation modes like $\Phi^{\dagger}_{DM} \Phi_{DM} \rightarrow \phi \phi , \phi' \phi', \phi \phi' , f \bar{f}$.

\begin{figure}[t!]
\centering
\includegraphics[angle=0,height=7.5cm,width=7.5cm]{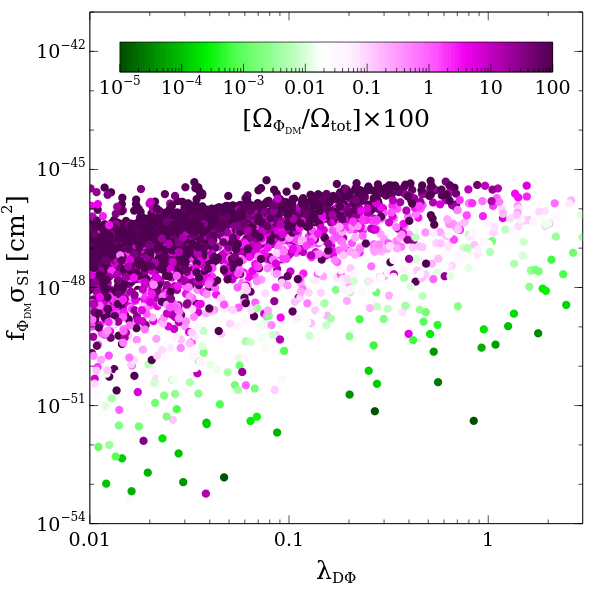}
\includegraphics[angle=0,height=7.5cm,width=7.5cm]{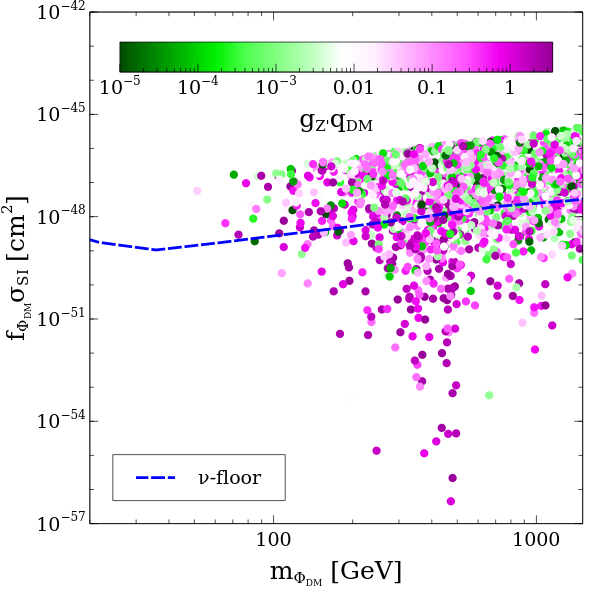}
\caption{left and right panels show the scatter plots in the $\lambda_{D\phi}-f_{\Phi_{DM}} \sigma_{SI}$ and $m_{\Phi_{DM}} - f_{\Phi_{DM}} \sigma_{SI}$ planes, respectively. The color variation in the left panel represents the amount of scalar DM in percentage compared to the total amount of DM density, whereas the right panel shows the variation of different values $g_{Z^{\prime}} q_{DM}$.} 
\label{DM-plot-5}
\end{figure}

In Fig.~\ref{DM-plot-5}, we show the scatter plots in the $\lambda_{D \phi}-f_{\Phi_{DM}} \sigma_{SI}$ and $m_{DM} - f_{\Phi_{DM}} \sigma_{SI}$ planes. The colorbar in the left panel describes the fraction of scalar DM present in percentage, while the right panel represents the different values of $g_{Z^{\prime}} q_{DM}$. In the left panel, it can be seen that if we increase the value of the quartic coupling $\lambda_{D\phi}$, then more annihilation occurs and there are more green points in the plot. It is also found that low values of effective direct detection spin-independent cross-section lead to a low abundance of scalar DM. This is evident from the colour variation in the left panel. On the other hand, in the right panel we show the scatter plot in the $m_{DM} - f_{\Phi_{DM}} \sigma_{SI}$ plane. In this case, all points are obtained after passing through the LUX-ZEPLIN direct detection bound. It can be seen in the right panel that the points around the $\nu-$floor have both higher and lower values of $g_{Z^{\prime}} q_{DM}$ depending on the values of the quartic couplings $\lambda_{D\phi},\,\, \lambda_{D\phi' }$. We can also see the points which correspond to the lower values of the SIDD cross-section (deep magenta points) represent a very small fraction of scalar DM. We can finally see there is a fraction in the allowed region below the $\nu-$floor which will be hard to probe by the current set-up of the direct detection experiments.

\section{Gravitational waves and dark matter}
\label{sec:GW_DM}

This section shows the correlations between GW and DM via the thermodynamic parameters of the PTs after implementing the bounds coming from DM relic density, collider and direct detection. As we will see in the plots a large portion of the parameter spaces are ruled out after implementing the collider and dark sector bounds. In showing the correlations, we have varied the model parameters in the range shown in Eqn.~(\ref{angle_range}). Moreover, we have kept $g_{Z^{\prime}}$ and $m_{Z^{\prime}}$ fixed at $10^{-3}$ and $0.1$~GeV, respectively, and require $|g_{Z^{\prime}} q_{DM}| \leq \sqrt{4 \pi}$.

\begin{figure}[t!]
\centering
\includegraphics[angle=0,height=7.0cm,width=7.0cm]{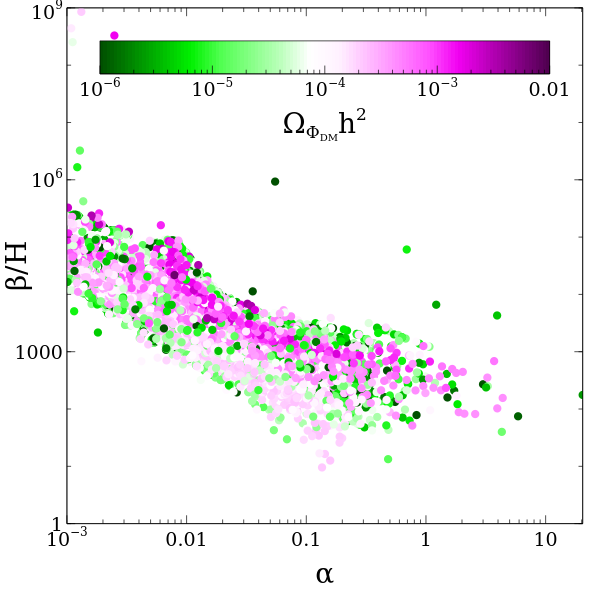}
\includegraphics[angle=0,height=7.0cm,width=7.0cm]{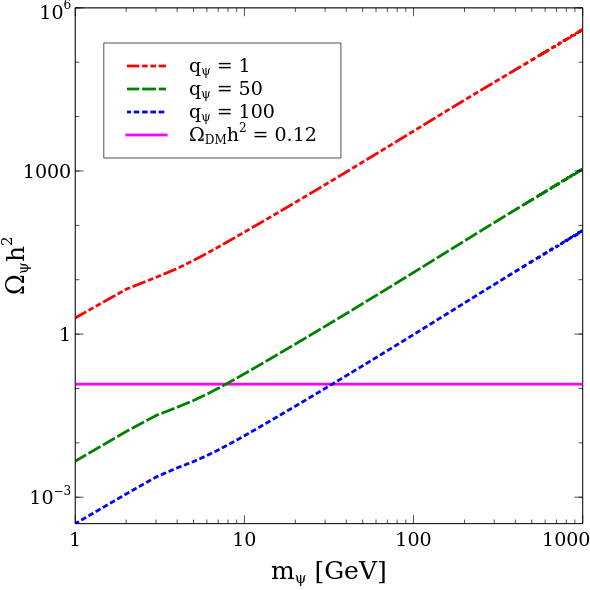}
\includegraphics[angle=0,height=7.0cm,width=7.0cm]{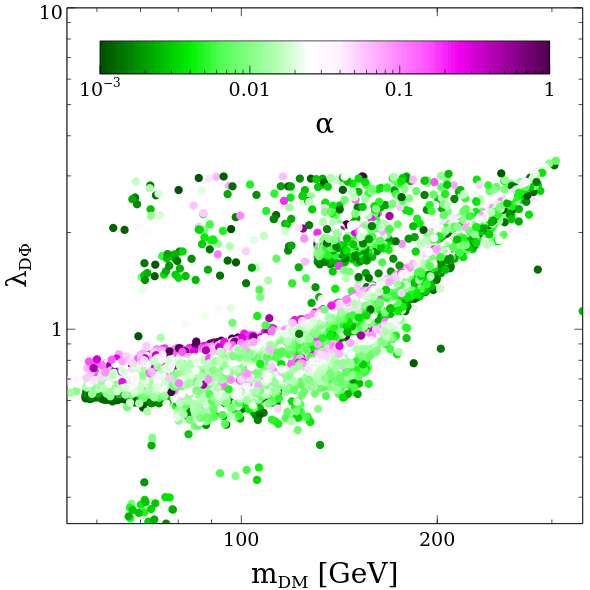}
\includegraphics[angle=0,height=7.0cm,width=7.0cm]{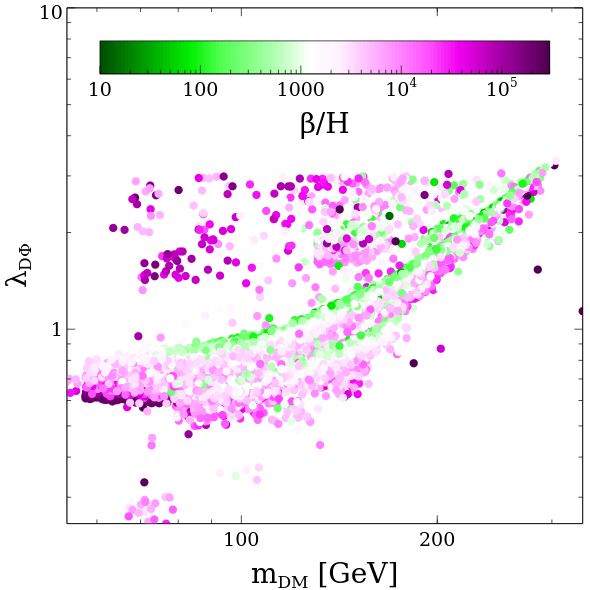}
\caption{Upper: Left panel shows the scatter plot in 
the $\alpha-\beta/H$ plane where the colour variation shows
the scalar DM relic density. Right panel shows the variation of the fermion DM relic density 
with its mass for three different values of $U(1)_{L_{\mu} - L_{\tau} }$ charges 
of the fermionic DM. Lower: let and right show the scatter plots in 
the $m_{DM}-\lambda_{D\phi}$ plane where the colour variation shows
different values of $\alpha$ ($\beta/H$) in the left panel (right panel).  }
\label{DM-GW-plot-1}
\end{figure}

In the left upper panel of Fig.~\ref{DM-GW-plot-1}, we have shown the scatter plot in the $\alpha-\beta/H$ plane where the colour variation implies the variation of scalar DM relic density. We can see anti-correlation between $\alpha$ and $\beta/H$. This is expected, as the larger values of $\alpha$ mean a larger separation between the two vacua, resulting in 
a larger transition time between the two vacua and vice versa. The larger values of $\alpha$
also mean higher emissions of latent heat during the transition which increases the Hubble 
parameter. In the plots, we have shown the points that give strong first-order PTs
along the $0,\,2,\,3$ directions as indicated in Table~\ref{PT_pattern}. As can be seen from the lower panels of Fig.~\ref{DM-GW-plot-1}, one needs higher values of $\lambda_{D\phi}$ 
to obtain the strong first-order PTs prompting more annihilations for the scalar DM. We get at most $\Omega_{\Phi_{DM}} h^{2} = 0.01$. This is not a problem because we also have fermionic 
vector DM in the present work which can give us the rest of the contribution in the total DM relic density. This is seen in the upper right panel plot where we have shown the variation of the fermionic DM relic density variation with its mass for three different values of the charge $q_{\psi}$. From the plot, we can see that as we increase the value of $q_{\psi}$ then we have a lower contribution from $\psi$ in the DM relic density for the same value $\psi$ mass, $m_{\psi}$. Therefore, by tuning the $q_{\psi}$ values we can easily satisfy the total DM relic density given by Planck. In the lower panel, the left and right plots show the allowed points in the $m_{DM} - \lambda_{D\phi }$ plane after demanding strong first-order PTs. The colour variations both in the left panel and right panel show the values of $\alpha$ and $\beta/H$. The values of these two parameters are anti-correlated, as exhibited in the colour variation. The lower panels of Fig.~\ref{DM-GW-plot-1} indicate that one needs high values of the quartic coupling $\lambda_{D\phi}$ to have the strong first-order PTs. These values of $\lambda_{D\phi}$ can also be constrained by the DM direct detection as will be shown later in this section.

\begin{figure}[t!]
\centering
\includegraphics[angle=0,height=7.5cm,width=6.5cm]{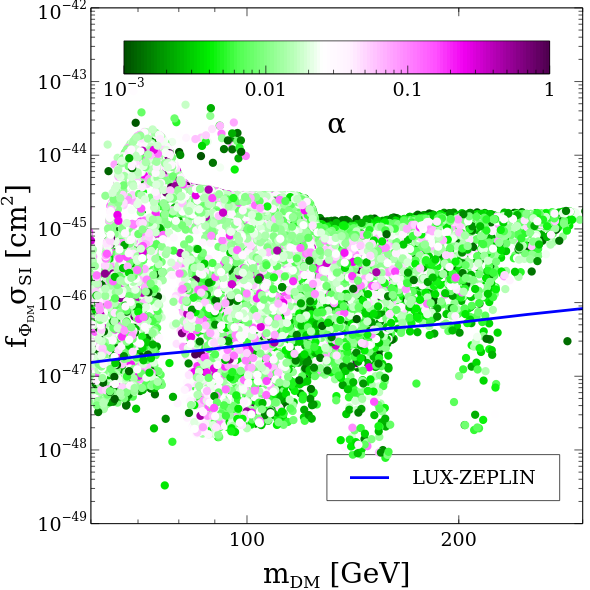}
\includegraphics[angle=0,height=7.5cm,width=6.5cm]{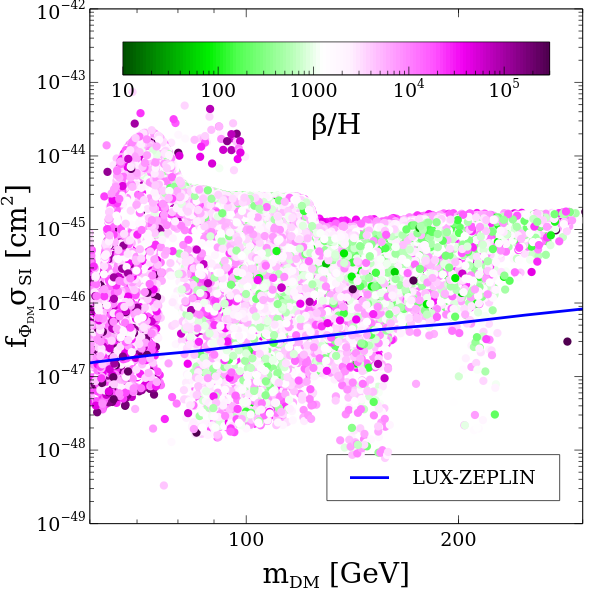}
\caption{ Scatter plots in the $m_{\Phi_{DM}} - f_{\Phi_{DM}} \sigma_{SI} $ plane where the colour bar in the left (right) shows the variation of $\alpha$ ($\beta/H$).} 
\label{DM-GW-plot-2}
\end{figure}

In Fig.~\ref{DM-GW-plot-2}, we show the scatter plots in the $m_{\Phi_{DM}}-f_{\Phi_{DM} \sigma_{SI}}$ plane after demanding the strong first-order PTs. In both panels, we can see through the colour variation the values of $\alpha, \beta/H$ distributed over the allowed region. As we discussed in context of the previous figure, one needs large values of $\lambda_{D\phi}$ to achieve strong first-order PTs, which results in a large portion of parameter space being ruled out by the direct detection (DD) experiment LUX-ZEPLIN. We will see in the next figure that the exclusion of a large portion of the parameter space by the DD does not hamper the potential of having all the allowed range of $\alpha$ and $\beta/H$ for the strong first-order PTs.  

\begin{figure}[t!]
\centering
\includegraphics[angle=0,height=7.5cm,width=6.5cm]{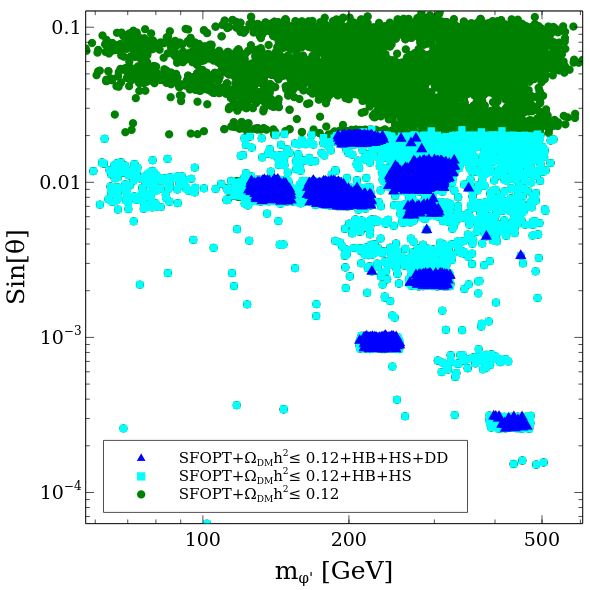}
\includegraphics[angle=0,height=7.5cm,width=6.5cm]{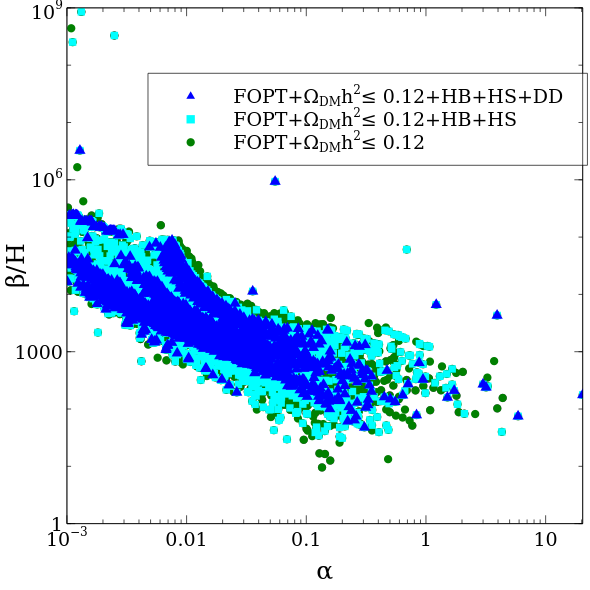}
\caption{ Left panel shows the scatter plot in the $m_{\phi' } - \sin \theta$ plane and RP shows the scatter plot in the $\alpha-\beta/H$ plane. Three different colours represent different bounds coming from strong first-order PTs, DM relic density, collider and direct detection on their subsequent implementation on the allowed data.} 
\label{DM-GW-plot-3}
\end{figure}

In the left panel and right panel of Fig.~\ref{DM-GW-plot-3}, we have shown scatter plots
in the $m_{\phi' }-\sin\theta$ and $\alpha-\beta/H$ planes after imposing a selection
of bounds from strong first-order PTs, DM relic density, collider and DM direct detection constraints. In both panels, the green points are obtained by implementing strong first-order PTs and upper bound on DM relic density {\it i.e.} $\Omega_{\Phi_{DM}} h^{2} \leq 0.12$, whereas the cyan points are obtained by additionally applying the collider bounds, which mainly come
from the Higgs signal strength and Higgs branching to invisible decay. Finally, the blue points are obtained by additionally applying the DM direct detection bound. In the case of collider bounds associated with the SM Higgs and additional Higgs have been implemented using the \texttt{HiggsSignal} and \texttt{HiggsBound}, respectively. In the left panel, we can see that when the collider bounds are applied, the values $\sin\theta > 0.02$ are
ruled out due to the SM Higgs decaying to the additional gauge boson and further decaying to neutrinos. In Ref.~\cite{ATLAS:2023tkt}, the ATLAS collaboration has displayed the allowed branching ratio for the SM Higgs invisible decay, $\text{Br}_{\phi \rightarrow \text{inv}} < 10.7\,\%$. Adopting this value, we have also checked analytically that,
\begin{eqnarray}
    \frac{\Gamma_{\phi \rightarrow Z^{\prime} Z^{\prime} }}{\Gamma_{\phi}} \leq 10.7 \%
    \Rightarrow \sin\theta < 0.02\,,
\end{eqnarray}
where $\Gamma_{\phi} = 4.1^{+5.0}_{-4.0} \,\,\text{MeV} $ \cite{CMS:2019ekd} and 
\begin{eqnarray}
    \Gamma_{\phi \rightarrow Z^{\prime} Z^{\prime} } = \frac{m^3_{\phi} \sin^{2}\theta }{32 \pi v'^2}
    \sqrt{1 - \frac{4 m^2_{Z^{\prime}}}{m^2_{\phi}}} \left( 1 - \frac{4 m^2_{Z^{\prime}}}{m^2_{\phi} }
    + \frac{12 m^4_{Z^{\prime}}}{m^4_{\phi} }
    \right)\,.
\end{eqnarray}
Moreover, when we apply the DM DD bound then we have fewer allowed spaces represented by the
blue points. Those blue points represent a very suppressed fraction of the scalar DM relic density in the range $\Omega_{\Phi_{DM}} h^{2} \simeq 10^{-5}-10^{-7}$, as otherwise the points would be ruled out by the DM SIDD bound. This can easily be understood from Eqns.~\eqref{si-dd-cs} and~\eqref{dm-dm-phi-phiP}, which imply that DM SIDD is linearly proportional to $\lambda_{D\phi}$ and suppression can only appear in the case the model has a very small fraction of the scalar DM, $f_{\Phi_{DM}}$. On the other hand, in the right panel, we can see that even after implementing the stronger bound from the DD, we can have relatively all the allowed range
for $\alpha$ and $\beta/H$ which we obtain after implementing the strong first-order PTs bound as well. Therefore, we may conclude that severe bounds from the collider and direct detection experiments can reduce the allowed parameter space but do not alter GW signal strength significantly.

\section{Conclusions}
\label{sec:conclusions}

The possibility of gauging the $U(1)_{L_{\mu}-L_{\tau}}$ symmetry as an extension of the Standard Model is a well-motivated endeavour and has recently become a subject of increasing interest. In this work, we investigated a non-minimal version of this model, featuring two complex singlet scalars, three Majorana right-handed neutrinos and a vector-like fermion. 

Like in its original version, the model offers a simple resolution to the long-standing puzzle of the muon $(g-2)$ anomaly via the presence of an additional one-loop diagram mediated by the $Z'$ gauge boson. At the same time, the RHNs provide a viable type-I seesaw mechanism, and the choice of $U(1)_{L_{\mu}-L_{\tau}}$ charge for one of the extra complex scalars induces a two-zero minor structure along the diagonal elements of the RHN neutrino mass matrix. We tested the viability of the neutrino oscillation parameters predicted in this model in light of the recent data from the NuFit collaboration and found that the model can successfully account for all experimental observables. Additionally, we confirmed that maximal $\theta_{23} = \pi/4$ as well as maximal CP phase $\delta_{CP}$ are possible in this model only for large values of the sum of neutrino masses, leading to a potential conflict with cosmological data. The present model can also explain the $1\,\sigma$ allowed range for $\theta_{23}$, with/without atmospheric data representing the lower/higher octant values without violating any other constraints. Moreover, we determined the $\delta_{CP}$ using the Jarlskog variable and found values within the allowed range provided by the NuFit collaboration.

It is well known that the interaction of an additional scalar with the SM Higgs doublet can easily trigger a potential barrier and render the EW phase transition strongly first-order. Thus, it is perhaps not surprising that our model, with its three-dimensional potential, offers a vast array of possibilities in terms of symmetry-breaking patterns. We studied the finite temperature evolution of the one-loop effective potential, taking into account all three field-space dimensions consisting of the SM Higgs doublet and two additional singlet scalars, including the DM candidate. Our analysis showed that the strongest phase transitions occur in a well-defined range of the DM squared-mass parameter corresponding to values $\mu_{DM}^2 \geq -v^2$ and that the Lagrangian parameter $\mu_{DM}^2$ has the strongest correlation with the phase transition latent heat. We have chosen $\mu_{DM}^2$ values such that we always get $m_{\phi_{DM}} > 0$ by
following Eqn.~\ref{mphidm}.

We then studied dark matter in association with GW and discussed the associated phenomenology. First, we focused on the new annihilation mode of scalar DM to the additional gauge bosons. This can be achieved by considering large values of the $U(1)_{L_{\mu} - L_{\tau}}$ gauge charge of the scalar DM. Small values of the charge can also serve the purpose if we do not aim to explain the muon $(g-2)$. This part of the DM study does not depend on the quartic couplings $\lambda_{D\phi}$ and $\lambda_{D\phi'}$, but this is not true when we focus on GW. Demanding the strong first-order PTs requires large values of the quartic couplings, which affect our scalar DM, from its abundance to the bound from direct detection. We found that the parameters associated with the scalar DM corresponding to the strong first-order PTs give us negligible contributions to the DM relic density, {\it i.e.}, $\Omega_{DM} h^{2} < 0.01$. Moreover, when we apply the DM direct detection bound on scalar DM, the scalar DM contribution further diminishes to $\Omega_{DM} h^{2} \sim 10^{-5}-10^{-7}$. Therefore, to obtain the remainder of DM contribution, we introduce another vector fermion DM, which can easily fill the rest of the DM density without altering the other phenomenologies. The additional fermion DM mainly annihilates to the additional gauge boson $Z'$, and its contribution fully depends on the 
$U(1)_{L_{\mu}-L_{\tau}}$ charge values taken for the fermion DM. We also found that $\sin\theta > 0.02$ is ruled out mainly due to the tight constraint on the SM Higgs invisible decay width.

In summary, our present model can simultaneously explain several beyond SM problems, namely muon $(g-2)$, neutrino mass, dark matter, and first-order phase transition. In the future, we will further explore the relationship between GW and baryon asymmetry in the canonical leptogenesis mechanism, as well as the Dirac CP phase in the present setup.


\section*{Acknowledgements}
The computations were enabled by resources provided by the National Academic Infrastructure for Supercomputing in Sweden (NAISS) at KTH partially funded by the Swedish Research Council through grant agreement no. 2022-06725.
This work used the Scientific Compute Cluster at GWDG, the joint data center of Max Planck
Society for the Advancement of Science (MPG) and University of G\"{o}ttingen. This project has received funding /support from the European Union's Horizon 2020 research and innovation programme under the Marie Sk\l{}odowska-Curie grant agreement No 860881-HIDDeN.

\appendix

\section{Neutrino parameters}
\label{app:neutrino_parameters}

The results obtained in this work have been checked against the \texttt{NuFit 5.2} dataset~\cite{Esteban:2020cvm}. In this section, the constrained values of the neutrino oscillation parameters are presented for the normal ordering of neutrino masses. When the atmospheric neutrino data from SuperKamiokande is taken into account, the best-fit values and associated 1$\sigma$ CL uncertainties are given by
\begin{equation}
    \sin^2{\theta_{12}} = 0.303_{-0.012}^{+0.012},
\end{equation}
\begin{equation}
\Delta m^2_{21} = 7.41_{-0.20}^{+0.21} \times 10^{-5} \ \text{eV}^2
\end{equation}
\begin{equation}
    \sin^2{\theta_{23}} = 0.451_{-0.016}^{+0.019},
\end{equation}
\begin{equation}
\Delta m^2_{31} = 2.507_{-0.027}^{+0.026} \times 10^{-3} \ \text{eV}^2
\end{equation}
\begin{equation}
    \sin^2{\theta_{13}} = 2.225_{-0.059}^{+0.056} \times 10^{-2},
\end{equation}
\begin{equation}
    \delta_{CP} = 232_{-26}^{+36} \ ^\circ.
\end{equation}
When the atmospheric neutrino data from SuperKamiokande is not taken into account, the best-fit values for the neutrino mixing parameters are instead
\begin{equation}
    \label{appeq_1}
    \sin^2{\theta_{12}} = 0.303_{-0.011}^{+0.012},
\end{equation}
\begin{equation}
    \label{appeq_2}
    \Delta m^2_{21} = 7.41_{-0.20}^{+0.21} \times 10^{-5} \ \text{eV}^2
\end{equation}
\begin{equation}
    \sin^2{\theta_{23}} = 0.572_{-0.023}^{+0.018},
\end{equation}
\begin{equation}
    \label{appeq_3}
    \Delta m^2_{31} = 2.511_{-0.027}^{+0.028} \times 10^{-3} \ \text{eV}^2
\end{equation}
\begin{equation}
    \label{appeq_4}
    \sin^2{\theta_{13}} = 2.203_{-0.059}^{+0.056} \times 10^{-2},
\end{equation}
\begin{equation}
    \delta_{CP} = 197_{-25}^{+42} \ ^\circ.
\end{equation}

The three mixing angles $\theta_{12}$, $\theta_{13}$ and $\theta_{23}$ can also be extracted from the PMNS matrix $U$ as follows~\cite{ParticleDataGroup:2018ovx}:
\begin{equation}
\sin^2{\theta_{13}} = |U_{13}|^2,
\end{equation}
\begin{equation}
\sin^2{\theta_{23}} =\frac{ |U_{23}|^2}{ 1 - |U_{13}|^2},
\end{equation}
\begin{equation}
\sin^2{\theta_{12}} = \frac{ |U_{12}|^2}{ 1 - |U_{13}|^2}.
\end{equation}

To calculate the {\em CP} phase $\delta_{CP}$, we use the Jarlskog invariant, which is given by
\begin{align}
J_\text{CP}
&\equiv \operatorname{Im}\big[ U_{\alpha i} U_{\alpha j}^* U_{\beta i}^* U_{\beta j} \big]
\\
&\equiv J_\text{CP}^\text{max} \sin \delta_{CP} =
c_{12} s_{12}
c_{23} s_{23} c^2_{13} s_{13}
\sin\delta_{CP} \,,
\label{jarlskog}
\end{align}
where $c_{ij}$ and $s_{ij}$ stand for $\cos \theta_{ij}$ and $\sin \theta_{ij}$, respectively. The $CP$ phase can therefore be solved from the equation
\begin{align}
\label{sin_delta}
    \sin \delta_{CP} = \frac{\operatorname{Im}\big[ U_{\alpha i} U_{\alpha j}^* U_{\beta i}^* U_{\beta j} \big]}{J_{CP}^\text{max}}.
\end{align}
In order to determine the correct geometrical phase for $\delta_{CP}$, one must also obtain an analytical expression for $\cos \delta_{CP}$. This can be achieved by examining the real part of the matrix product $\operatorname{Re}\big[ U_{\alpha i} U_{\alpha j}^* U_{\beta i}^* U_{\beta j} \big]$, which results in
\begin{align}
\label{cos_delta}
    \cos \delta_{CP} = \frac{\operatorname{Re}\big[ U_{\alpha i} U_{\alpha j}^* U_{\beta i}^* U_{\beta j} \big] + s_{12}^2 s_{13}^2 c_{13}^2 s_{13}^2}{J_{CP}^\text{max}}.
\end{align}
The $CP$ phase $\delta_{CP}$ can now be determined by comparing the values of $\sin \delta_{CP}$ and $\cos \delta_{CP}$ in Eqns.~(\ref{sin_delta}) and~(\ref{cos_delta}). If $\sin \delta_{CP} > 0$ and $\cos \delta_{CP} > 0$, $\delta_{CP}$ can be obtained directly from $\sin \delta_{CP}$. If $\sin \delta_{CP} > 0$ and $\cos \delta_{CP} < 0$, the correct $CP$ phase is obtained with transformation $\delta_{CP} \rightarrow \pi - \delta_{CP}$. If $\sin \delta_{CP} < 0$ and $\cos \delta_{CP} < 0$, the phase is given as $\delta_{CP} \rightarrow \delta_{CP} + \pi$ instead. Finally, if $\sin \delta_{CP} < 0$ and $\cos \delta_{CP} > 0$, then $\delta_{CP} \rightarrow - \delta_{CP}$.

The charged lepton flavor violation constraints from MEG~\cite{MEGII:2023ltw} are taken into account by calculating the elements of the active-sterile neutrino mixing matrix $U_{\nu N}$. Following Ref.~\cite{Morisi:2024yxi}, we obtain the mixing between the light and heavy neutrinos from $M_D$ and $M_R$ matrices: 
\begin{align}
    \label{Ue2}
    U_e^2 = \sum_i \left| \left(U_{\nu N}\right)_{e i} \right|^2 = \sum_i \left(m_D M_R^{-1}\right).
\end{align}
We take the MEG limit on the radiative muon decay, $Br(\mu \rightarrow) \lesssim 7.5\times10^{-13}$, and calculate the upper bound for $U_e$. Any data points not satisfying the MEG limit are then excluded in the numerical analysis, as discussed in section~\ref{sec:neutrinos}. The model parameters values obtained in the neutrino analysis fall within the following ranges:
\begin{equation}
    M_{ee} \in [-3.4,88.3] \times 10^{7}, \nonumber
\end{equation}
\begin{equation}
    M_{\mu\tau} \in [-1.2,167.0] \times 10^{6}, \nonumber
\end{equation}
\begin{equation}
    \xi \in [-7.3,58.5] \times 10^{4}, \nonumber
\end{equation}
\begin{equation}
    h_{e\mu} \in [-1.3,34.7] \times 10^{9}, \nonumber
\end{equation}
\begin{equation}
    h_{e\tau} \in [-1.7,9.2] \times 10^{9}, \nonumber
\end{equation}
\begin{equation}
    y_{e} \in [-10.5,7.6] \times 10^{-4}, \nonumber
\end{equation}
\begin{equation}
    y_{\mu} \in [-15.6,9.5], \nonumber
\end{equation}
\begin{equation}
    y_{\tau} \in [-1.1,3.8].
\end{equation}
All of the data points satisfy the experimental constraints specified in Eqns.~\eqref{appeq_1}-\eqref{appeq_4}. It should be noted that while some of the values considered for $y_\mu$ and $y_\tau$ are large, the majority of the values obtained for these two parameters are close to zero. It is also found that the values presented for $\xi$ are multiples of $\pi$, which indicates that the off-diagonal terms $M_{\mu\tau} e^{i\xi}$ in $M_R$ are always real.

\section{Field dependent masses}

At finite temperature, all the radial modes from each scalar can mix and the mass matrix can be written as
\begin{align}
 M^2 & \equiv \frac{\partial^2 V_0}{\partial{\phi_i}\partial{\phi_j} } \nonumber \\
  &= \begin{pmatrix}
      \scriptstyle \frac{1}{2} (\chi^2 \lambda_{D\phi} + 6 \phi^2 \lambda_{\phi} + \phi'^2 \lambda_{\phi \phi'} - 2 \mu_{\phi}^2)   &     \scriptstyle \phi \phi' \lambda_{\phi \phi'} &       \scriptstyle \phi \chi \lambda_{D\phi}  \\
     \scriptstyle \phi \phi' \lambda_{\phi\phi'}   &      \scriptstyle \frac{1}{2} (\chi^2 \lambda_{D\phi'} + 6 \phi'^2 \lambda_{\phi'} + \phi^2 \lambda_{h\phi'}) + \mu_{\phi'}^2 &  \scriptstyle \phi' \chi \lambda_{D\phi'} \\
    \scriptstyle \phi \chi \lambda_{D\phi} & \scriptstyle \phi' \chi \lambda_{D\phi'} &  \scriptstyle \frac{1}{2} (\phi^2 \lambda_{D\phi} + \phi'^2 \lambda_{D\phi'} + 6 \chi^2 \lambda_{DM}) + \mu_{DM}^2 \\
    \end{pmatrix} \label{massmat_scalars}
\end{align}

The field dependent masses for the scalars are obtained from the eigenvalues of Eqn.~\eqref{massmat_scalars}. Since closed-form expressions are difficult to obtain or they are very lengthy, we choose not to write them down here. Instead, they are calculated numerically. 

On the other hand, the field dependent masses of the gauge bosons are easy to obtain, namely
\begin{equation}
m_W^2(\phi)  =\frac{g^2}{4}\phi^2, \quad m_Z^2(\phi)=\frac{g^2+g'^2}{4}\phi^2, \quad m_{Z'}^2(\phi',\chi) = g_Z^2(q_{DM}^2 \chi^2 + \phi'^2).
\end{equation}

The contribution from the quark sector is dominated by the top quark since it has the largest Yukawa coupling, {\it i.e.},
\begin{equation}
  m_t^2(\phi)=\frac{y_t^2}{2}\phi^2 \, .
\end{equation}
We neglect the contribution from other quarks in this work. 

The case of the right-handed neutrinos requires more attention as the field dependent mass matrix is given by~\cite{Biswas:2016yan}
\begin{eqnarray}
\mathcal{M}_{R}(\phi') = \left(\begin{array}{ccc}
M_{ee} ~~&~~ \dfrac{ \phi'}{\sqrt{2}} h_{e \mu}
~~&~~\dfrac{\phi' }{\sqrt{2}} h_{e \tau} \\
~~&~~\\
\dfrac{\phi'}{\sqrt{2}} h_{e \mu} ~~&~~ 0
~~&~~ M_{\mu \tau} \,e^{i\xi}\\
~~&~~\\
\dfrac{\phi'}{\sqrt{2}} h_{e \tau} ~~&
~~ M_{\mu \tau}\,e^{i\xi} ~~&~~ 0 \\
\end{array}\right) \,.
\label{mncomplex}
\end{eqnarray} 
One can multiply the matrix in Eqn.~\eqref{mncomplex} by its Hermitian conjugate and digonalize the result. Therefore,
\begin{equation}
    U^{\dagger} M_R^{\dagger}(\phi')M_R(\phi') U = \text{diag} \left( m_{N_1}^2(\phi'),\ m_{N_2}^2(\phi'),\ m_{N_3}^2(\phi') \right),
\end{equation}
where each neutrino eigenstate has $4$ degrees of freedom. We do not include the contribution from the left-handed neutrinos as their masses are tiny.

\subsection{Daisy resummation}
The thermally-corrected masses of the scalars are the eigenvalues of 
the thermally-corrected Hessian matrix
\begin{equation}
M^2 \rightarrow M^2 +
    \begin{pmatrix}
    \Pi_{\phi}(T) &  &\\
     &  \Pi_{\phi'}(T)&\\ 
    &&\Pi_{\chi}(T)
    \end{pmatrix} \, .
\end{equation}
This corresponds to a resummation of a large number of daisy diagrams which enhance the perturbative control at large temperatures~\cite{Parwani:1991gq,Curtin:2016urg}. 

The thermal masses in this model are found to be 
\begin{equation}
 \Pi_{\phi} = \left( \frac{3g^2}{16} + \frac{g'^2}{16} + \frac{\lambda_{\phi}}{2} + \frac{\lambda_{\phi \phi'}}{12} + \frac{\lambda_{D\phi}}{12}  + \frac{y_t^2}{4} \right) T^2 \, ,
\end{equation}
\begin{equation}
\Pi_{\phi'} =\left(  \frac{g_Z^2 }{4  }    +  \frac{\lambda_{\phi'}}{3}  + \frac{\lambda_{\phi \phi'}}{6} + \frac{\lambda_{D\phi'}}{12}  +  \frac{{h_{e\mu}^2 + h_{e\tau}^2}}{12} \right)T^2
\end{equation}
and
\begin{equation}
\Pi_{\chi} =\left(  \frac{g_Z^2  q_{DM}^2}{4  }    +   \frac{\lambda_{DM}}{3}  + \frac{\lambda_{D\phi}}{6} + \frac{\lambda_{D\phi'}}{12}  \right)T^2 \, .
\end{equation}

The longitudinal polarization states of vector bosons receive temperature corrections as
\begin{equation}
    m_{\text{gauge}}^2 \rightarrow m_{\text{gauge}}^2 + \Pi_{\text{gauge}}(T).
\end{equation}
For the SM $W$ bosons, this reads~\cite{Katz:2014bha}
\begin{equation}
\Pi_{W} = \frac{11}{6}g^2 T^2
\end{equation}
and the eigenvalues of the mass matrix
\begin{equation}
M_{Z/\gamma}^2 =
    \begin{pmatrix}
    \frac{1}{4}g^2 \phi^2 + \frac{11}{6}g^2 T^2 & -\frac{1}{4}g g' \phi^2\\
    -\frac{1}{4}g g' \phi^2 & \frac{1}{4}g'^2 \phi^2 + \frac{11}{6}g'^2 T^2\\
    \end{pmatrix}
\end{equation}
determines the finite temperature mass of the $Z$ boson and the photon. Similarly, the longitudinal component of the $Z'$ acquires the correction
\begin{equation}
    \Pi_{Z'} = \frac{1}{6} g_Z^2\left( q_{DM}^2  + 4 \right)T^2.
\end{equation}

\bibliographystyle{JHEP}
\bibliography{references.bib}  
\end{document}